\documentclass[twocolumn,secnumarabic,amssymb,nobibnotes,aps,prd,floatfix,superscriptaddress]{revtex4-2}

\usepackage{graphicx}
\usepackage{dcolumn}
\usepackage{bm}

\RequirePackage{graphicx}
\usepackage{xcolor}
\usepackage{subfig}
\usepackage{booktabs}
\usepackage{amsmath}
\usepackage{amssymb}
\usepackage{relsize}
\usepackage{makecell}
\usepackage{multirow}
\usepackage{mhchem}
\usepackage{caption}

\usepackage{floatrow}
\newfloatcommand{capbtabbox}{table}[][\FBwidth]

\usepackage{letltxmacro}
\LetLtxMacro{\oldcite}{\cite}
\renewcommand{\cite}[1]{\mbox{\oldcite{#1}}}

\begin{document}

\preprint{APS/123-QED}

\title{Enhanced low-energy supernova burst detection in large liquid argon time projection chambers enabled by Q-Pix}

\collaboration{The Q-Pix Collaboration}

\newcommand{\Harvard}{Department of Physics, Harvard University, Cambridge, MA 02138, USA}
\newcommand{\Arlington}{Department of Physics, University of Texas at Arlington, Arlington, TX 76019, USA}
\newcommand{\Manchester}{University of Manchester, Manchester M13 9PL, UK} 
\newcommand{\Wellesley}{Physics Department, Wellesley College, Wellesley, MA 02481, USA}
\newcommand{\Fermilab}{Fermi National Accelerator Laboratory, Batavia, IL 60510, USA}
\newcommand{\LBNL}{Lawrence Berkeley National Laboratory, Berkeley, CA 94720, USA}
\newcommand{\ORNL}{Oak Ridge National Laboratory, Oak Ridge, TN 37831, USA}
\newcommand{\ANL}{Argonne National Laboratory, Lemont, IL 60439, USA}
\newcommand{\Hawaii}{Department of Physics and Astronomy, University of Hawaii, Honolulu, HI 96822, USA}
\newcommand{\UPenn}{Department of Physics and Astronomy, University of Pennsylvania, Philadelphia, PA 19104, USA}

\author{S.~Kubota}
\affiliation{\Harvard}
\author{J.~Ho}
\thanks{Corresponding author: \url{johnnyho@fas.harvard.edu}}
\affiliation{\Harvard}
\author{A.D.~McDonald}
\thanks{Corresponding author: \url{austin.mcdonald@uta.edu}}
\affiliation{\Arlington}
\affiliation{\Harvard}
\author{N.~Tata}
\affiliation{\Harvard}
\author{J.~Asaadi}
\affiliation{\Arlington}
\author{R.~Guenette}
\affiliation{\Manchester}
\affiliation{\Harvard}
\author{J.B.R.~Battat} 
\affiliation{\Wellesley}
\author{D.~Braga} 
\affiliation{\Fermilab}
\author{M.~Demarteau} 
\affiliation{\ORNL}
\author{Z.~Djurcic}
\affiliation{\ANL}
\author{M.~Febbraro} 
\affiliation{\ORNL}
\author{E.~Gramellini} 
\affiliation{\Fermilab}
\author{S.~Kohani}    
\affiliation{\Hawaii}
\author{C.~Mauger}
\affiliation{\UPenn}
\author{Y.~Mei}  
\affiliation{\LBNL}
\author{F.M.~Newcomer}
\affiliation{\UPenn}
\author{K.~Nishimura}
\affiliation{\Hawaii}
\author{D.~Nygren}
\affiliation{\Arlington}
\author{R.~Van~Berg} 
\affiliation{\UPenn}
\author{G.S.~Varner}  
\affiliation{\Hawaii}
\author{K.~Woodworth} 
\affiliation{\Fermilab}

\date{\today}

\begin{abstract}

The detection of  neutrinos from core-collapse supernovae may reveal important process features  as well as neutrino properties. The detection of supernova neutrinos is one of the main science drivers for future kiloton-scale neutrino detectors based on liquid argon.  Here we show that for such detectors the intrinsically 3D readout in Q-Pix  offers numerous advantages relative to a wire-based readout, such as higher reconstruction efficiency, lower energy threshold, considerably lower data rates, and potential pointing information.

\end{abstract}

\maketitle

\section{Introduction}\label{sec:intro}
The central role neutrinos play in the fate of massive stars as they reach the end of their life in a core-collapse supernova (SN) has long been noted in the field of astrophysics~\cite{1966ApJ143626C}. Observing neutrinos emitted from a supernova provides new ways to test our understanding of both nuclear and particle physics at the most extreme densities and energies~\cite{1987ApJ322795F,Duan:2009cd,Lunardini:2015afa}. A core-collapsing star is a unique laboratory within which the dynamics of the death of a star as well as neutrino--neutrino interactions and oscillations may be observed in a way that cannot be reproduced with a terrestrial experiment.

In the emergent era of multi-messenger astronomy, technological advancements in neutrino detection are needed to allow a detailed, high-statistics description of the neutrino burst from a supernova collapse. Thus far, the direct detection of neutrinos from SN~1987A by three underground experiments~\cite{Kamiokande-II:1987idp, PhysRevLett.58.1494, 1987ApJ322795F} confirmed some aspects of supernova astrophysics and provided insight into how the detection of neutrinos from supernovae could lead to a deeper understanding of fundamental neutrino physics. 

The dense surroundings of a supernova is the only environment in the universe where neutrino flavor oscillations can be enhanced by the entirety of neutrino interaction phenomenon~\cite{SupernovaeFlavorOscillation} (e.g., multiple neutrino species and energies impacted by the Mikheyev--Smirnov--Wolfenstein effect, neutrino--neutrino interactions, and vacuum oscillations) as the neutrinos propagate through the multiple layers of stellar materials of wildly varying densities and types of interactions on their way to interstellar space~\cite{Duan:2009cd}. Through observations of the energy and timing profiles of those events, fundamental neutrino and astrophysical parameters could be extracted, such as neutrino mass ordering~\cite{massOrdering}, neutrino halo model characteristics~\cite{haloModel}, and the neutrino magnetic moment~\cite{neutrinoMagnticMoment}. The $\mathcal{O}(10)$ events that were experimentally observed from SN~1987A have already provided crucial information, but future core-collapse supernovae, plausibly anticipated within the next few decades, may provide even greater detail. For these reasons, next-generation neutrino experiments must be prepared to accurately measure the energy, timing structure, and flavor of the neutrino spectrum~\cite{JUNO:2015zny, Hyper-Kamiokande:2018ofw, DUNE:2020zfm, Hyper-Kamiokande:2021frf, JUNO:2021vlw}. 

Large-scale noble element time projection chambers (TPCs)~\cite{Nygren:PEP:1974} play a central role in many aspects of high-energy physics, both currently running and planned in the near future. Charged particles traversing the bulk  material produce ionization electrons and scintillation photons. An imposed electric field forces the ionization electrons to drift toward the detector anode where they are collected on charge-sensitive readout. The combined measurement of the scintillation light, providing the $t_0$, with the arrival time of the ionization charge, allows for the 3D reconstruction of the original charged particle topology. Thus, the TPC provides a fully active and uniform tracking detector with calorimetric reconstruction capabilities without instrumenting the bulk volume of the detector.

The capability to drift electrons over many meters has made the use of large-scale liquid noble TPCs attractive as neutrino detectors to study neutrino oscillations over relatively short ($<$1~km)~\cite{Machado:2019oxb} and long baselines ($>$1000~km)~\cite{acciarri2016long}. Specifically, liquid argon time projection chambers (LArTPCs)~\cite{willis1974liquid,rubbia1977liquid} offer fine-grained tracking as well as powerful calorimetry and particle identification capabilities. This makes LArTPCs ideal detectors for studying neutrino--nucleus and neutrino--electron interactions as well as neutrino oscillation phenomena. A review of the recent experimental applications of LArTPCs is given in Ref.~\cite{Majumdar:2021llu}. 

A conventional method for reading out the ionization charge in a LArTPC relies on the use of consecutive planes of sensing wires to measure two of the three spatial coordinates using the 2D projections to reconstruct the 3D image. This method was used for ICARUS~\cite{antonello2015operation} and MicroBooNE~\cite{acciarri2017design}, as well as many other recent experiments~\cite{Anderson:2012vc,LArIAT:2019kzd,Bian:2015qka}. This configuration was also adopted as a baseline configuration for the Deep Underground Neutrino Experiment (DUNE) far detector~\cite{acciarri2016long}. Although the concept is proven and has gained considerable support in the community, it has an intrinsic limitation in resolving ambiguities in dense, complex topology reconstruction making the event reconstruction difficult in some cases. Novel event reconstruction techniques may be employed to overcome these difficulties~\cite{Qian:2018qbv, MicroBooNE:2020vry, MicroBooNE:2020jgj, MicroBooNE:2021zul, MicroBooNE:2021ojx}, however, reconstruction using 3D-based pixel readouts can offer significant improvements~\cite{Adams:2019uqx}. Moreover, the long sense wires used in large-scale LArTPCs introduce significant capacitance to the readout electronics~\cite{MicroBooNE:2017qiu} which may limit the extraction of physics signals at low-energy thresholds [e.g., $\mathcal{O}(\leq$MeV)]~\cite{LArIAT:2019gdz}. These detectors also produce large volumes of raw data as every wire is read  out continuously [$\mathcal{O}($ms)]. Finally, the construction and mounting of massive anode plane assemblies to host the wires pose difficult engineering challenges and can be quite expensive. For these reasons, a native 3D, fully-independent pixel readout could provide advantages.

A pixel-based readout scheme has been utilized in small-scale gas TPCs~\cite{Gonzalez-Diaz:2017gxo} but had not been previously considered for very large LArTPCs because of the much larger number of readout channels, and the high data rate and power consumption.  A transformative step forward for future LArTPCs would be the ability to build a fully pixelated low-power charge readout capable of efficiently and accurately capturing signal information. The potential for scientific gain through the realization of a low-power, pixel-based charge readout for use in LArTPCs has independently inspired two research groups to pursue complimentary approaches to solving this problem. The LArPix~\cite{Dwyer:2018phu} and Q-Pix~\cite{Nygren:2018rbl} consortia have undertaken the needed R\&D to realize such a readout.

The Q-Pix solution, discussed in more detail in Section~\ref{sec:Q-PixOverview} and in Ref.~\cite{Nygren:2018rbl}, targets the daunting requirements established for the DUNE far detectors~\cite{DUNETDR}. These detectors must be capable of high efficiency ${\nu}_{\mu} / \bar{\nu}_{\mu}$ and ${\nu}_{e} / \bar{\nu}_{e}$ discrimination and precise energy reconstruction to provide a definitive measurement of the $CP$-phase and to identify the neutrino mass hierarchy. Additionally, these detectors must be capable of detecting low-energy neutrinos originating from supernovae bursts and have the ability to detect the signature of baryon number violation (via proton decay or neutron/anti-neutron oscillations)~\cite{Abi:2018dnh}. These far detector modules will only record $\sim$4 beam events/day/10~kton module with mean neutrino energies $\mathcal{O}$(GeV) while having to be simultaneously sensitive to much rarer and significantly lower energy events (such as those that come from a supernova burst) which will give $\mathcal{O}(100\textrm{-}1000)$ events (depending on the distance of the supernova) in a short time window [$<$$\mathcal{O}(10)$~seconds].  A description and first study of supernova burst signals in the DUNE far detector are provided in Ref.~\cite{DUNE:2020zfm}.

In this paper we  further elaborate low-energy  reconstruction in the Q-Pix scenario for supernova events.  Section~\ref{sec:Q-PixOverview} provides an overview of the Q-Pix architecture. Section~\ref{sec:LowEPhysSim} describes the simulation tools used to model the supernova neutrino interactions, the simulation of the detector and of the radiogenic backgrounds, and the simulation of the Q-Pix readout. Finally, Section~\ref{sec:Results} presents the results focusing on Q-Pix's ability to reconstruct low-energy ($<$5~MeV) events with high efficiency and purity in the presence of radiogenic backgrounds, the trigger efficiency from a ``charge-only'' readout, the data rates expected from a Q-Pix module, and finally the ability to do directional pointing using supernova neutrino events reconstructed in the simulation.

\begin{figure*}[t]
  \begin{center}
    \includegraphics[width=0.65\textwidth]{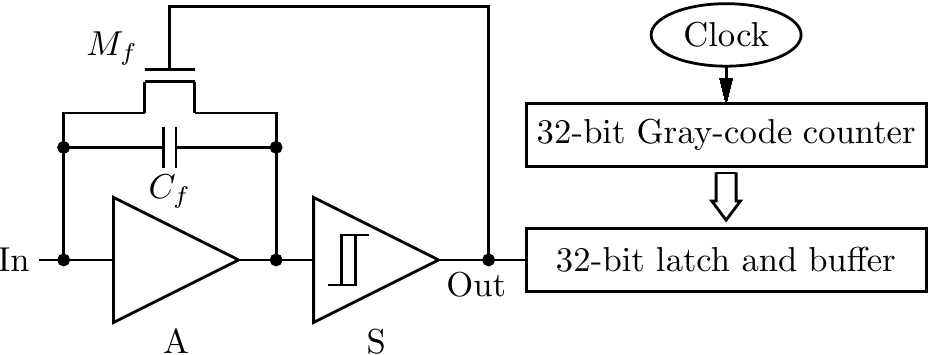}
    \caption{Representation of the charge integrator/reset circuit where a free-running oscillator increments a Gray-code counter in order to create a local clock. A $reset$ signal causes the local clock to be stored in a buffer register. The data is a string of clock snapshots, from which a Reset Time Difference (RTD) can be calculated.}
    \label{fig:charge-integrator-reset}
  \end{center}
\end{figure*}

\begin{figure*}[t]
  \begin{center}
    \includegraphics[width=0.45\textwidth]{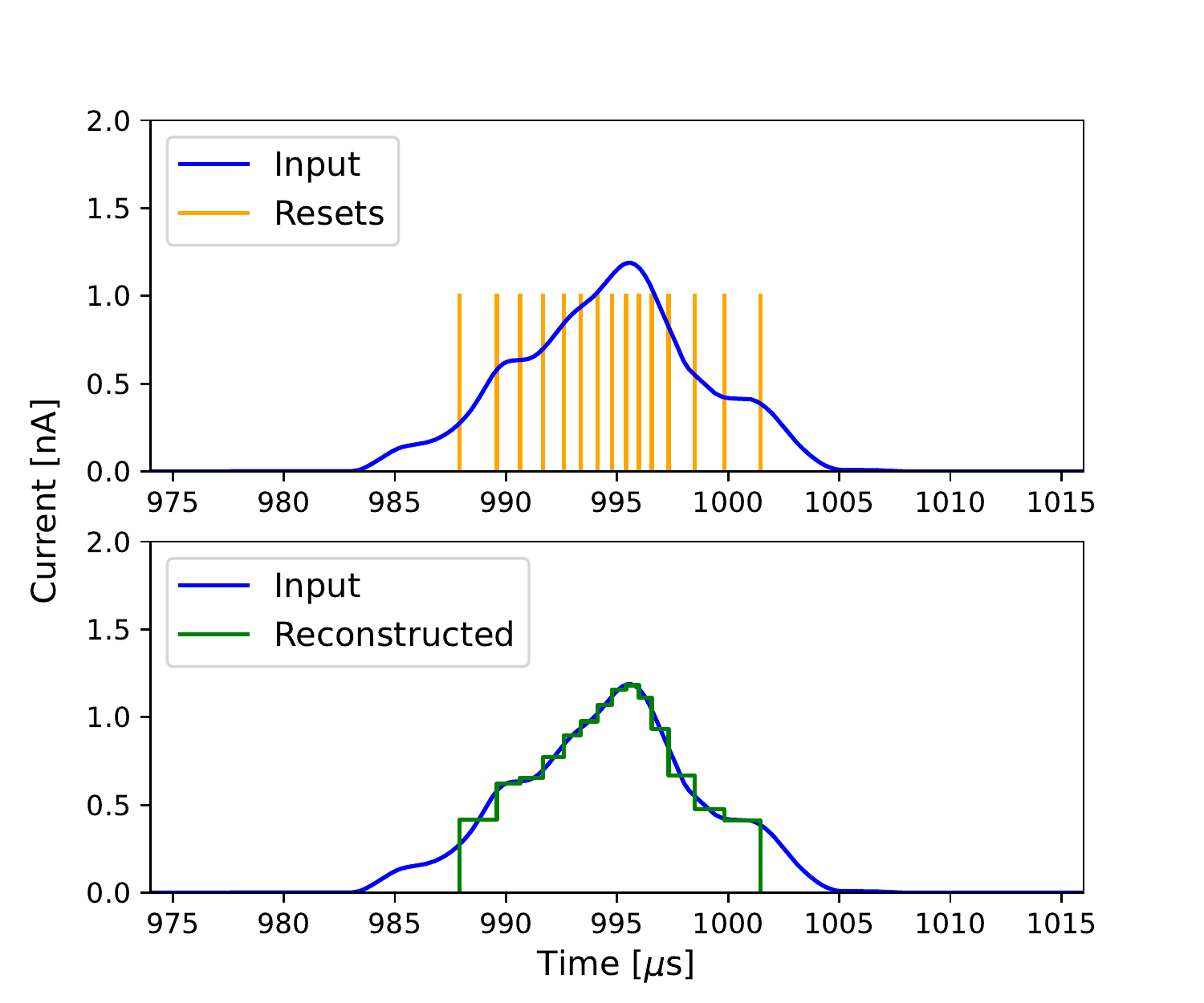}
    \includegraphics[width=0.45\textwidth]{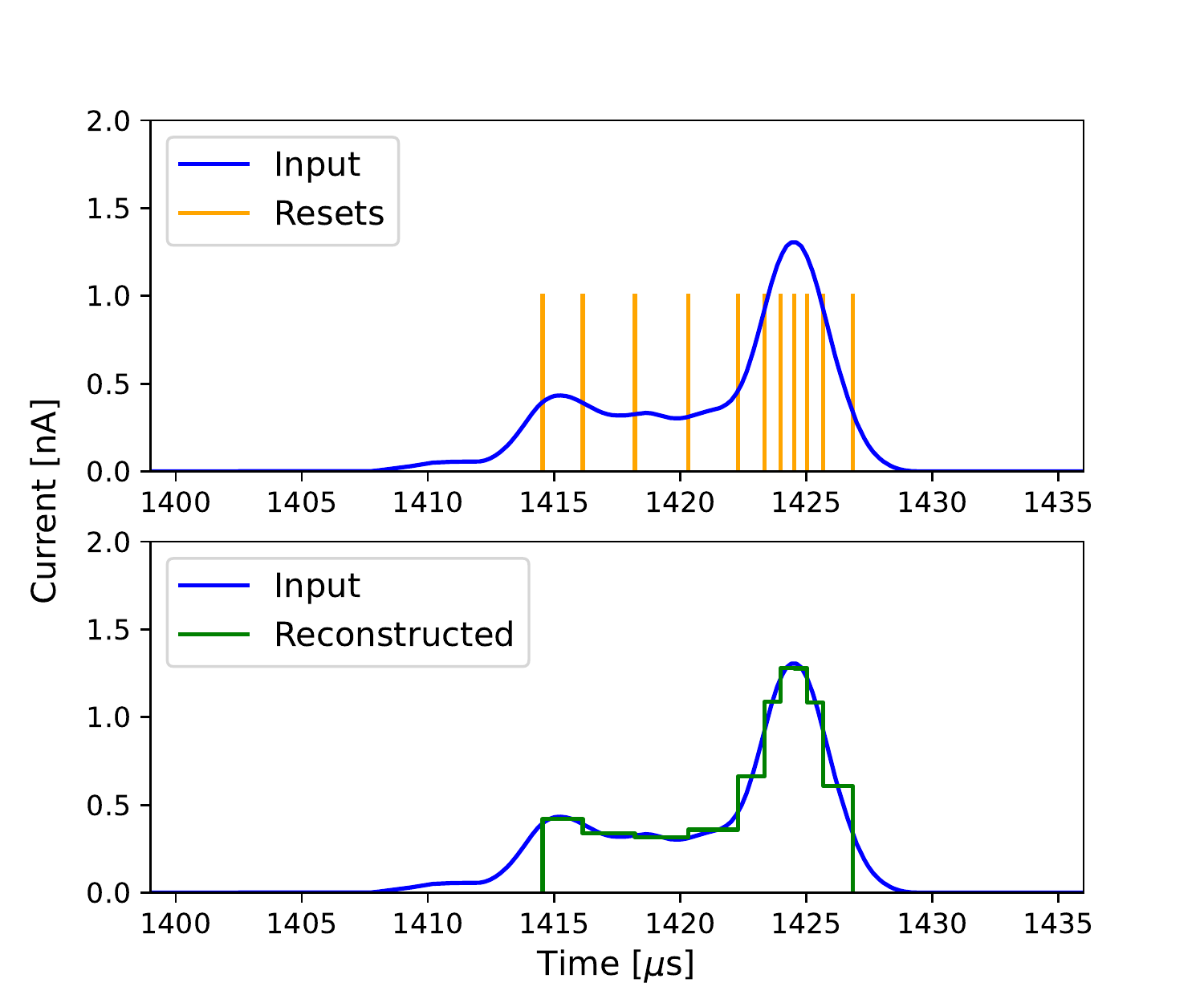}
    \caption{Charge integration simulation results for two different pixels near the vertex of a simulated neutrino interaction. The current arriving at the pixel is shown in blue; the corresponding resets are shown in orange and the reconstructed current from the Reset Time Differences (RTDs) is shown in green. Signal waveforms can be reconstructed from RTDs since $I \propto 1/\mathrm{RTD}$ where $I$ is the average current over an interval $\Delta T$ and thus $I \cdot \Delta T = \int I(t) \,\mathrm{d}t = \Delta Q$. The $\Delta Q$ for the charge-integration/reset is fixed and was chosen to be 1~fC.}
    \label{fig:QPixImage}
  \end{center}
\end{figure*}

\section{Overview of the Q-Pix readout}\label{sec:Q-PixOverview}
The fundamental idea of the Q-Pix readout scheme is to use pixel-scale self-triggering `charge integrate/reset' blocks with free-running clocks and dynamically established data networks robust against single-point failure (SPF)~\cite{Nygren:2018rbl}. This pixelization concept is targeted as a `technology of opportunity' for a multi-kiloton DUNE far detector (FD). In the DUNE FD, high-precision data across all spatial and energy ranges is desired for signal events, but most of the time, nothing of interest is occurring in the detector and the data acquisition scheme should be idle until something happens. The ethos of ``do not do anything unless there is something to do'' can be thought of as an electronic principal of least action and is a design philosophy at the heart of the Q-Pix architecture. This design idea provides a solution to the needed low-power architecture to operate in a single-phase (SP) LArTPC as well as simultaneously solving the large data rates which come with a high-granularity readout.  

The basic concepts of the Q-Pix circuit is shown in Fig.~\ref{fig:charge-integrator-reset}. The input pixel, labeled as ``In,'' is envisioned to be a simple circular trace connected to the Q-Pix circuit by a via in a printed circuit board. The circuit begins with the \textbf{Charge-Integrate/Reset} (CIR) circuit. This charge sensitive amplifier continuously integrates incoming signals on a feedback capacitor until a threshold on a Schmitt trigger (regenerative comparator) is met. When this threshold is met, the Schmitt trigger starts a rapid ``reset'' which drains the feedback capacitor and returns the circuit to a stable baseline and the cycle is free to begin again. This ``reset'' transition pulse is used to capture and store the present time of a local clock within one ASIC (application-specific integrated circuit). This changes the basic quantum of information for each pixel from the traditional ``charge per unit of time'' data format to the difference between one clock capture and the next sequential capture, referred to as the \textbf{Reset Time Difference} (\textbf{RTD}). This new unit of information measures the time to integrate a pre-defined charge ($\Delta Q$). Physics signals will produce a sequence of short [$\mathcal{O}(\mu s)$] RTDs. In the absence of a signal, the quiescent input current from backgrounds (\ce{^{39}Ar}, cosmogenic, and other radioactivity) would be small and the expected RTDs are on the order of seconds.

Signal waveforms can be reconstructed from RTDs by exploiting the fact that the average input current and the RTD are inversely correlated ($I \propto 1/\mathrm{RTD}$), where $I$ is the average current over an interval $\Delta T$ and thus $I \cdot \Delta T = \int I(t) \,\mathrm{d}t = \Delta Q$. The signal current is captured with fixed $\Delta Q$, determined by the charge integrator/reset circuit, but with varying time intervals. An initial study of the requirements for the minimum $\Delta Q$ for the Q-Pix circuit ($\sim$6000 electrons)  as well as the range and precision of $\Delta T$ has been carried out using simulated signals from neutrino interactions fed into a full simulation of the Q-Pix circuit in the TSMC 180 nm process design kit (PDK), however, the ultimate limit of how low in threshold this technology can achieve is yet to be determined and is the focus of an upcoming prototype and subsequent paper. For the purposes of this paper, we have assumed conservative and realizable benchmarks for both the minimum $\Delta Q$ and precision of $\Delta T$, summarized in Section \ref{sec:Q-PixSimulation}. Fig.~\ref{fig:QPixImage} shows two examples of the conversion from RTDs back to arbitrary charge input of a particular $\Delta Q$ of one femtocoulomb (6250~electrons, or equivalently, 147.5~keV in LAr). 

The time-stamping architecture currently envisioned for the Q-Pix readout will utilize a technique first pioneered by the IceCube Neutrino Observatory~\cite{Aartsen:2016nxy} and shown to work with high reliability and precision. A local clock based on a free-running oscillator within the ASIC is used and its value captured in a buffer register when a ``reset'' transition occurs. The string of ``reset'' times are transmitted periodically out of the cryostat and a linear transformation from local clock frequency to central master clock allows one to recover the universal time RTDs. The interrogation of the local clock by surface systems need only occur as necessary to monitor and correct for oscillator drift. For nominal electron drift speeds in liquid argon of 1.6~mm/$\mu$s, a global timing accuracy of $\pm 1~{\mu}s$ corresponding to $\sim$1.6~mm in the drift direction can be easily obtained. 

\begin{figure}[t]
    \centering
    \includegraphics[width=\textwidth]{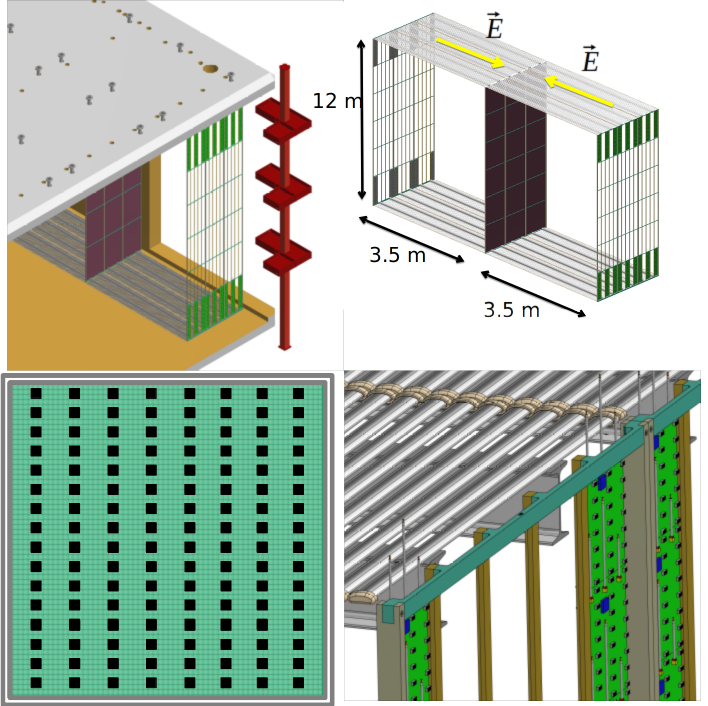}
    \caption{A conceptual representation of what a 10-kton DUNE far detector module using the Q-Pix tile readout. The top left figure shows a cutaway of the cryostat (in red) with a single TPC. The top right shows the dimensions of a TPC with the cathode in the center and the anode plane assemblies (APAs) with the Q-Pix tiles held in place. The tile boards can be housed in a frame similar to that used for the high voltage cathode and deployed in the existing design for the DUNE cryostat with minimal modifications. The lower left figure shows a conceptual drawing of the Q-Pix tile with each black square representing a single ASIC. The lower right shows a zoomed-in view of a single APA with tiles mounted.}
    \label{fig:QPixDetectorConcept}
\end{figure}

While the Q-Pix ASIC chip itself represents the smallest unit for the system, a more useful architecture which is resilient against single-point failure is to define a \textbf{\textit{tile}} as an array of $N \times N$ Q-Pix ASIC chips. As an illustrative example, if each Q-Pix ASIC chip, shown as black squares in the lower left of Fig.~\ref{fig:QPixDetectorConcept}, were to serve 16 pixels (in a matrix of ${4 \times 4}$ pixels), then this would constitute an array of a ${64 \times 64 = 4096}$ pixel block per tile, as the fundamental unit of the system. The number of Q-Pix ASICs, $N^2$, per tile and exact dimensions of the tile will be determined by the pixel pitch and number of channels per ASIC. These quantities themselves need to be studied to ensure the maximum physics reach for the given readout design, but will likely result in a tile size $\mathcal{O}(625$~cm$^2$) with pixel pitch of 4~$\times$~4~mm$^2$. These are the dimensions used in the study presented here. 

The design of the tile and the network between the ``$N^2$'' ASICs will be built following the design principle to make it intrinsically fault-tolerant and robust to as many possible failure modes. Each ASIC on the tile will have signal sensing, self-triggering, local clock, time-stamping, buffering, input/output, and state machine capabilities. Local time capture and data transfer could occur along any of the corner ASICs, offering a robust system design protective against SPF. The layout of the architecture suggests that 8 bits will be required to specify the ASIC position within the tile. 16 bits are needed to specify which individual pixel initiated the reset. For a clock running at 50~MHz, a 32-bit timestamp gives 43 seconds before wrap-around would occur (deemed more than sufficient for LArTPC operation) and leaves 8 bits yet unspecified to be used for other purposes. Preliminary studies suggest the power consumption of such a readout is quite low ($\sim$50~$\mu$W/channel). The quiescent data rate for a kiloton scale detector would also be low as discussed further in Section~\ref{sec:DataRates}.

The envisioned 10-kton module, which would serve as home to the Q-Pix readout, is similar by design to the ``conventional'' DUNE horizontal drift~\cite{DUNETDR} targeted as the first module and is illustrated in Fig.~\ref{fig:QPixDetectorConcept}. The idea is to simply substitute the existing anode plane assembly (APA) which houses the wire-based projective readout with a slightly modified cathode plane assembly (CPA) capable of hosting Q-Pix tiles. The most significant change is the central-most APAs which in the conventional DUNE design serve two drift regions simultaneously via a wrapped wire design. Instead, these central APAs will have a pair of ``back-to-back'' pixel planes oriented such that each drift region is independent. Thus the total number of APA drift regions for a 10-kton Q-Pix design will be 200 (instead of the 150 wire-based APAs).

However, it may very well turn out that a doubling of drift length is feasible, given technical advances in LAr purification. Drift lengths of 6.5 meters are already envisaged for the recently proposed vertical drift approach~\cite{Paulucci:2021sqn}. In this case, the central plane would be a single cathode plane, and the number of Q-Pix tiles and modified CPA assemblies would be reduced by a factor of two, with concomitant savings in cost and system complexity. 

For the analysis presented in this study, no photon detection is used and all the analysis is done using the collected ionization charge. The exact photon detection scheme to be used for such a multi-kiloton pixel-based readout is an area of active ongoing R\&D and thus is omitted from further discussion. Instead, only where noted, we assume that whatever photon system which is ultimately used will be able to provide $t_0$ for the events of interest.

\section{Low-Energy Physics Simulation} \label{sec:LowEPhysSim}
One area of particular interest to the multi-kiloton scale liquid argon experiments is the physics which can be enabled through exploration of low-energy ($<$100~MeV) phenomena~\cite{Castiglioni:2020tsu, Caratelli:2022llt}. Measurements of MeV-scale activity from accelerator neutrino interactions~\cite{ArgoNeuT:2018tvi}, radiogenic backgrounds~\cite{osti_1573057, MicroBooNE:2022his}, and Michel electrons~\cite{ICARUS:2003zvt, MicroBooNE:2017kvv, LArIAT:2019gdz} have already been performed by the ArgoNeuT, LArIAT, MicroBooNE, and ICARUS experiments. Low-energy LArTPC signatures from a variety of possible sources have been investigated and described in literature, including those from solar neutrinos~\cite{Capozzi:2018dat, Para:2022gju},  accelerator neutrinos~\cite{Friedland:2018vry, Castiglioni:2020tsu}, hidden sector~\cite{Harnik:2019zee, ArgoNeuT:2019ckq} particle and WIMP~\cite{Avasthi:2022tjr} interactions, neutrinoless double beta decay~\cite{Mastbaum:2022rhw}, and supernova neutrinos~\cite{DUNE:2020zfm, Castiglioni:2020tsu}, which will be the physics focus of this study. In this section we present the simulation framework used to quantify the increased physics reach of a DUNE LArTPC with Q-Pix readout relative to an APA readout.

Section~\ref{sec:SupernovaPhysics} provides an overview of the event generator and the supernova neutrino model utilized in this analysis. Section~\ref{sec:Backgrounds} presents the radiogenic backgrounds in the detector and the techniques utilized to integrate them into the simulation. Section~\ref{sec:Q-PixSimulation} provides details on the architecture of the Q-Pix readout simulation that was implemented.


\subsection{Supernova Neutrinos}\label{sec:SupernovaPhysics}
\begin{figure}[ht]
    \centering
    \includegraphics[width=\textwidth]{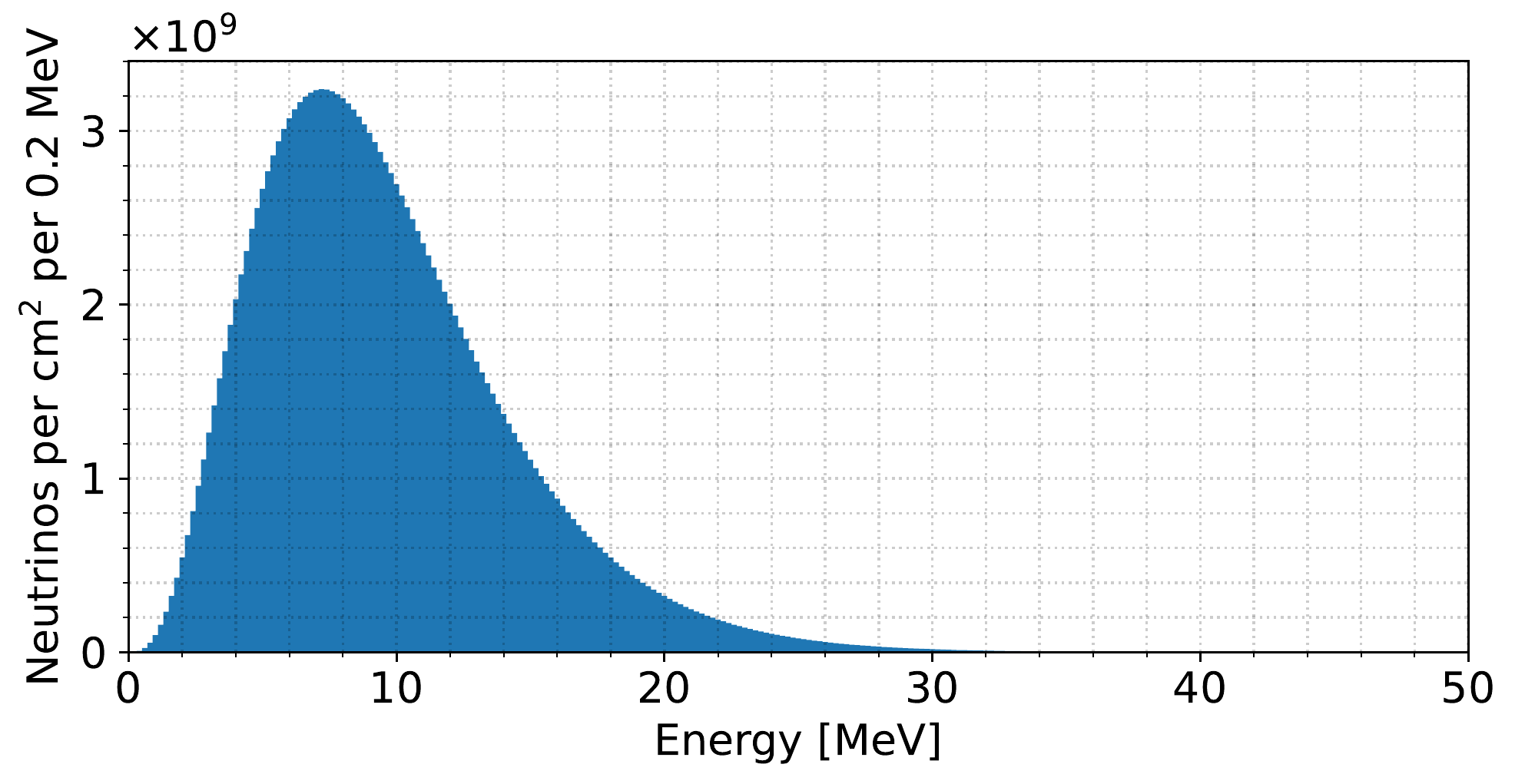}
    \includegraphics[width=\textwidth]{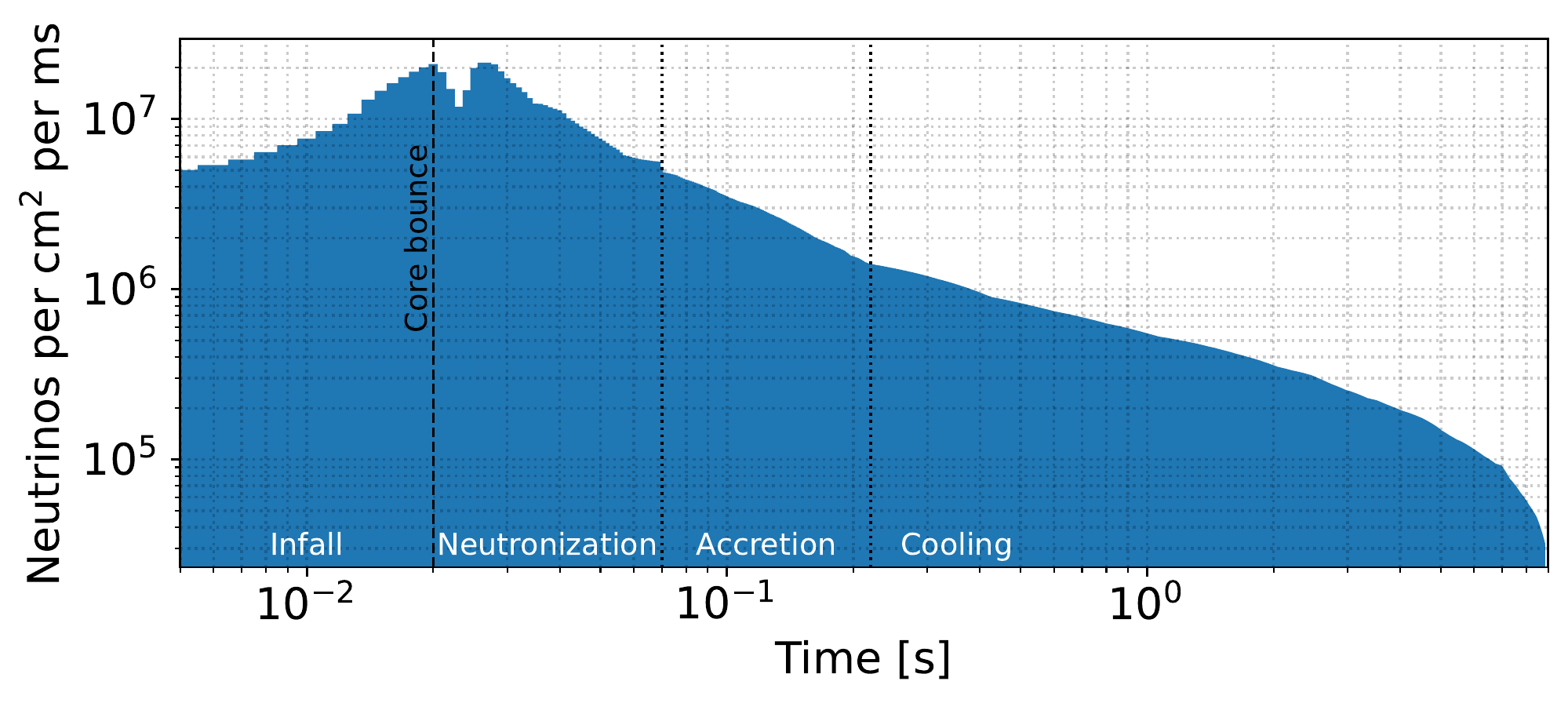}
    \includegraphics[width=\textwidth]{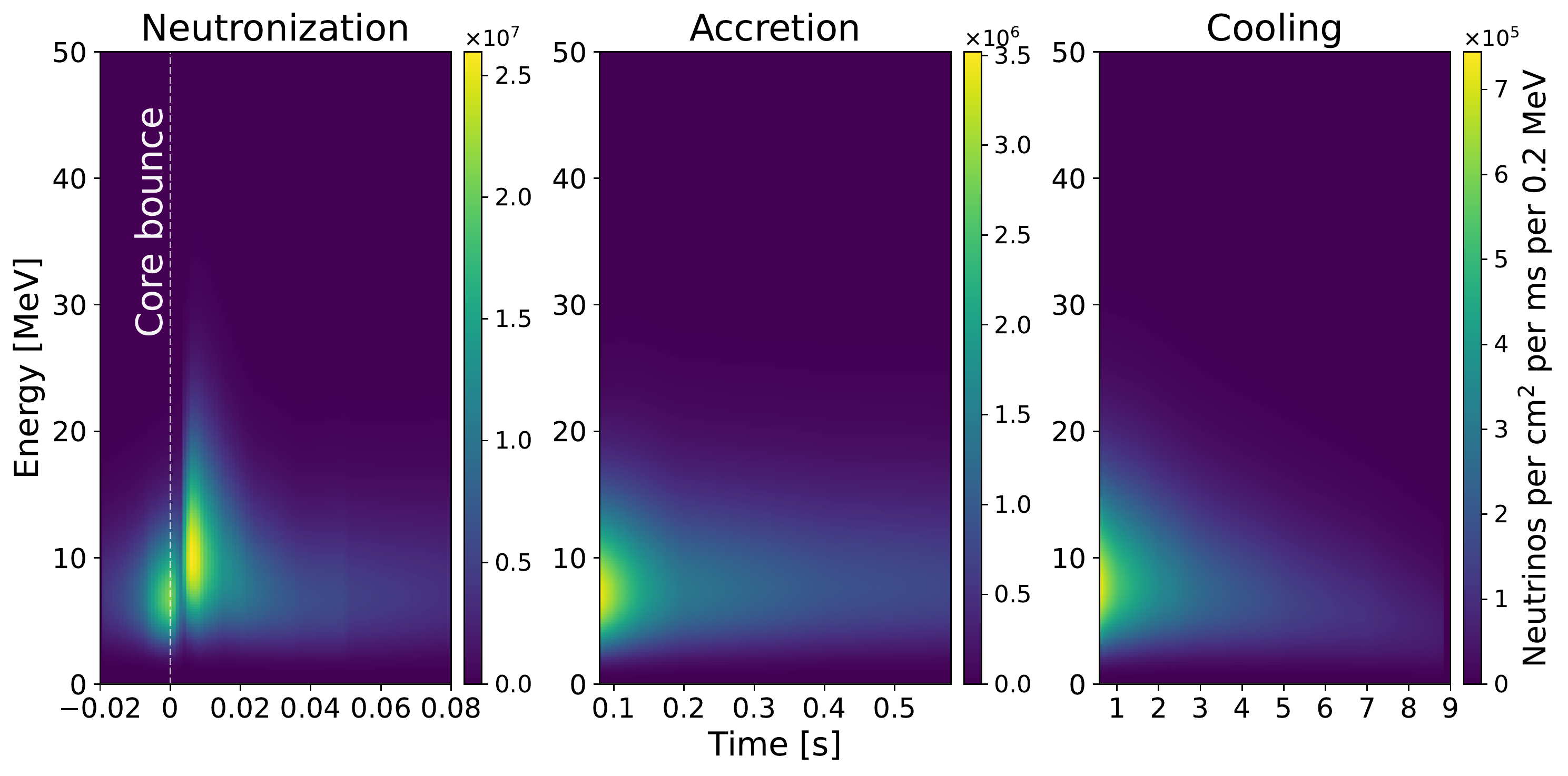}
    \caption{Top: the supernova $\nu_e$ energy spectrum of the ``Garching'' electron-capture supernova model used as an input to the {\tt MARLEY} event generator. Middle: the supernova $\nu_e$ timing profile indicated with different eras in the supernova evolution. Bottom: the time-dependent $\nu_e$ flux of the supernova separated into different eras in the supernova evolution. Adapted from Ref.~\cite{DUNE:2020zfm}.}
    \label{fig:SupernovaNovaNeutrinoEnergy}
\end{figure}

Our simulation of supernova neutrino events is based on supernova luminosity, energy, and time distributions provided by the ``Garching'' electron-capture supernova model~\cite{2010PhRvL.104y1101H} as propagated through {\tt SNOwGLoBES}~\cite{2011APSAPRL11006S} to provide event rates per 10-kton LAr module. This approach is similar to that of Ref.~\cite{DUNE:2020zfm}. This benchmark model was chosen as a pessimistic case as it predicts the fewest number of neutrino interactions per 10-kton detector. As has been done in other works~\cite{Suwa:2019svl, Lang:2016zhv, Odrzywolek:2010je, Seadrow:2018ftp}, we assume a supernova distance of 10~kiloparsecs (kpc) which predicts 220 charged-current (CC) electron neutrino ($\nu_e$) and 19 electron neutrino--electron (${\nu_e\text{--}e^{-}}$) elastic scattering (ES) interactions within a 10-kton module (Table~\ref{tab:snb-event-counts}). The energy and timing profiles of the simulated supernova neutrinos are shown in Fig.~\ref{fig:SupernovaNovaNeutrinoEnergy} and serve as inputs into our simulation.

For the simulation of the supernova neutrino--nucleus interactions, the {\tt MARLEY} (Model of Argon Reaction Low Energy Yields) event generator~\cite{gardiner2015marley,gardiner2021simulating} is used. The latest version of {\tt MARLEY} (v1.2.0) utilized in this study includes both $\nu_e$~CC and ${\nu_e\text{--}e^{-}}$~ES events. Thus the primary channels analyzed in this analysis are
\begin{align}
  \nu_e + \ce{^{40}Ar} &\rightarrow e^- + \ce{^{40}K^*} \tag{$\nu_e$ CC} \\
  \nu_e + e^- &\rightarrow \nu_e + e^- \tag{${\nu_e\text{--}e^{-}}$ ES}
\end{align}
where {\tt MARLEY} handles the subsequent nuclear de-excitation and produces a list of final-state observable particles. The electron anti-neutrinos ($\bar{\nu}_e$) and neutral-current interactions of any neutrino flavors (${\nu_X + \ce{Ar} \rightarrow \nu_X + \ce{Ar}^*}$) are not taken into account in this analysis.

\begin{table*}[th]
    \centering
    \setlength{\tabcolsep}{10pt}
    \def\arraystretch{1.0}%
    \resizebox{0.875\textwidth}{!}{%
    \begin{tabular}{ccccccc}
        \toprule
        Isotope       & Rate {[}Bq/kg{]} & Region            & Region mass {[}kg{]} & Rate {[}Bq{]} & \makecell{Number of decays \\ (per 10 s window)} \\
        \midrule
        \ce{^{210}Po} & 0.2              & PD {[}Bq/m$^2${]} & 2.46856              & 0.493712      & 5                \\
        \ce{^{60}Co}  & 0.0455           & CPA               & 90                   & 4.095         & 41               \\
        \ce{^{40}K}   & 0.49             & APA               & 258                  & 1,264.2       & 12,642           \\
        \ce{^{39}Ar}  & 1.010            & bulk LAr          & $\sim$70,000         & 70,700        & 707,000          \\
        \ce{^{42}Ar}  & 0.000092         & bulk LAr          & $\sim$70,000         & 6.44          & 64               \\
        \ce{^{42}K}   & 0.000092         & bulk LAr          & $\sim$70,000         & 6.44          & 64               \\
        \ce{^{222}Rn} & 0.04             & bulk LAr          & $\sim$70,000         & 2,800         & 28,000           \\
        \ce{^{214}Pb} & 0.01             & bulk LAr          & $\sim$70,000         & 700           & 7,000            \\
        \ce{^{214}Bi} & 0.01             & bulk LAr          & $\sim$70,000         & 700           & 7,000            \\
        \ce{^{85}Kr}  & 0.115            & bulk LAr          & $\sim$70,000         & 8,050         & 80,500           \\
        \bottomrule
    \end{tabular}
    }
    \caption{Summary of the radiogenic backgrounds, adapted from Ref.~\cite{shi2019studies}, outlining the particular radioactive isotope, the region the isotope originates from, the estimated decay rate for the isotope, and the expected number of decays in a 10-second simulation window.}
    \label{tab:radiogenic-backgrounds}
\end{table*}

\subsection{Backgrounds}\label{sec:Backgrounds}
An important part of the analysis of supernova burst detection capabilities is the inclusion of background events from radiogenic sources within the detector and the bulk argon. A previous analysis did not include these backgrounds~\cite{DUNE:2020zfm} since their selected signals required a minimum of 5~MeV of deposited energy within the argon~\cite{DUNE:2020ypp, DUNE:2020zfm} and thus would exclude the majority of the radiogenic backgrounds. To explore the capabilities of detection at lower energies, it is critical to include an estimate of these backgrounds. An initial estimate of the expected radiogenic isotopes and their expected activity levels was taken from Ref.~\cite{shi2019studies} with exact dimensions of the structures described in the detector taken from Ref.~\cite{DUNETDR}. This information is summarized in Table~\ref{tab:radiogenic-backgrounds}.

The typical timing profile of the neutrinos from a supernova burst spans $\sim$10~seconds and thus the radiogenic backgrounds are calculated over a full 10-second window. To appropriately handle the backgrounds and their associated rates, the decays associated with the radiogenic backgrounds are simulated as a single ``event'' over a 10-second window. Thus the single ``event'' contains $\sim$842,000 primary radioactive decays and are randomly distributed uniformly in time over that 10-second window. The radioactive isotopes are in equilibrium at time $t=0$ seconds to account for previous decays that could have occurred prior to the start of the readout.

\begin{figure*}[ht!]
     \centering
     \subfloat{{\includegraphics[width=0.47\textwidth]{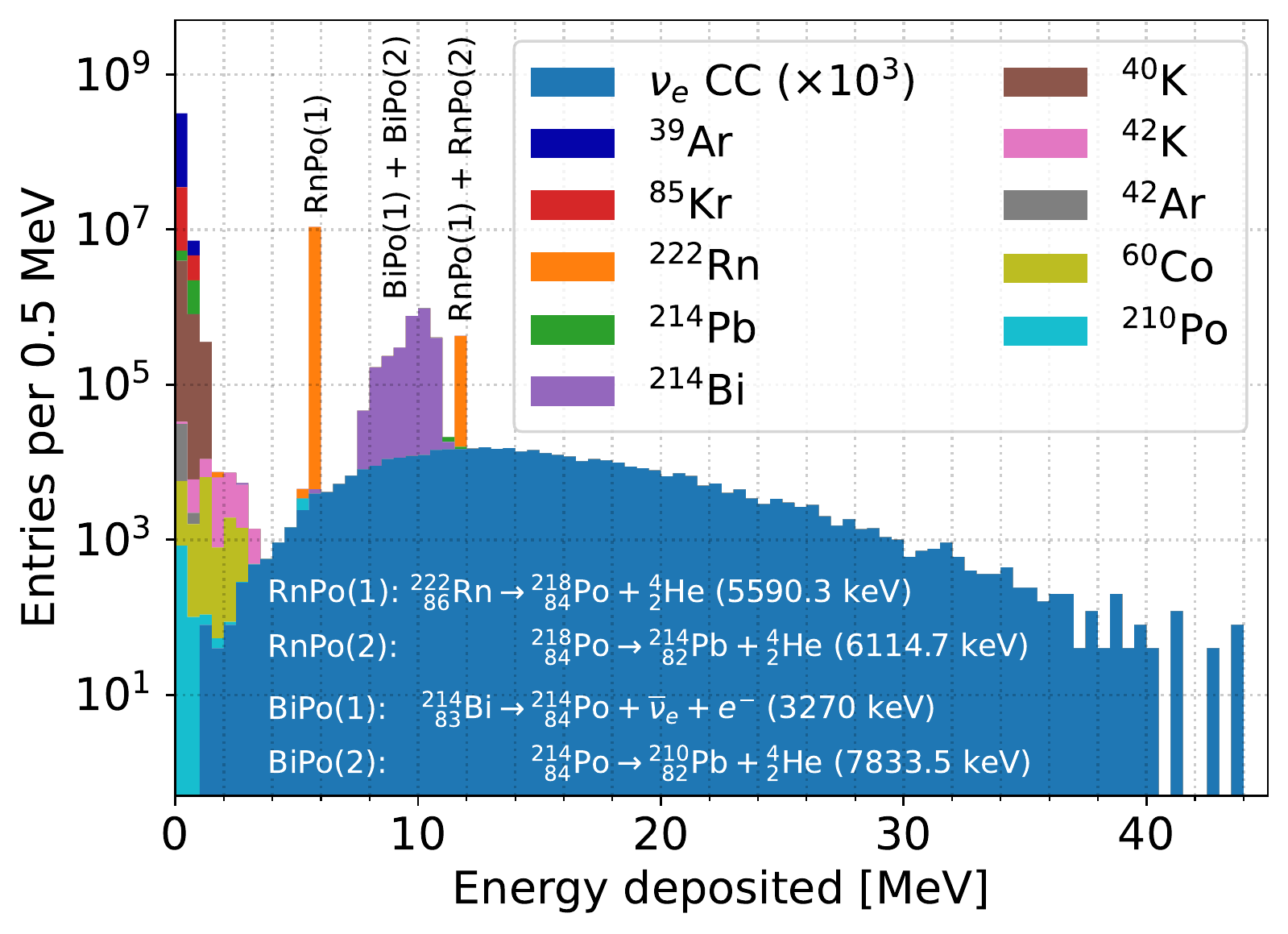}}}%
     \qquad
     \subfloat{{\includegraphics[width=0.47\textwidth]{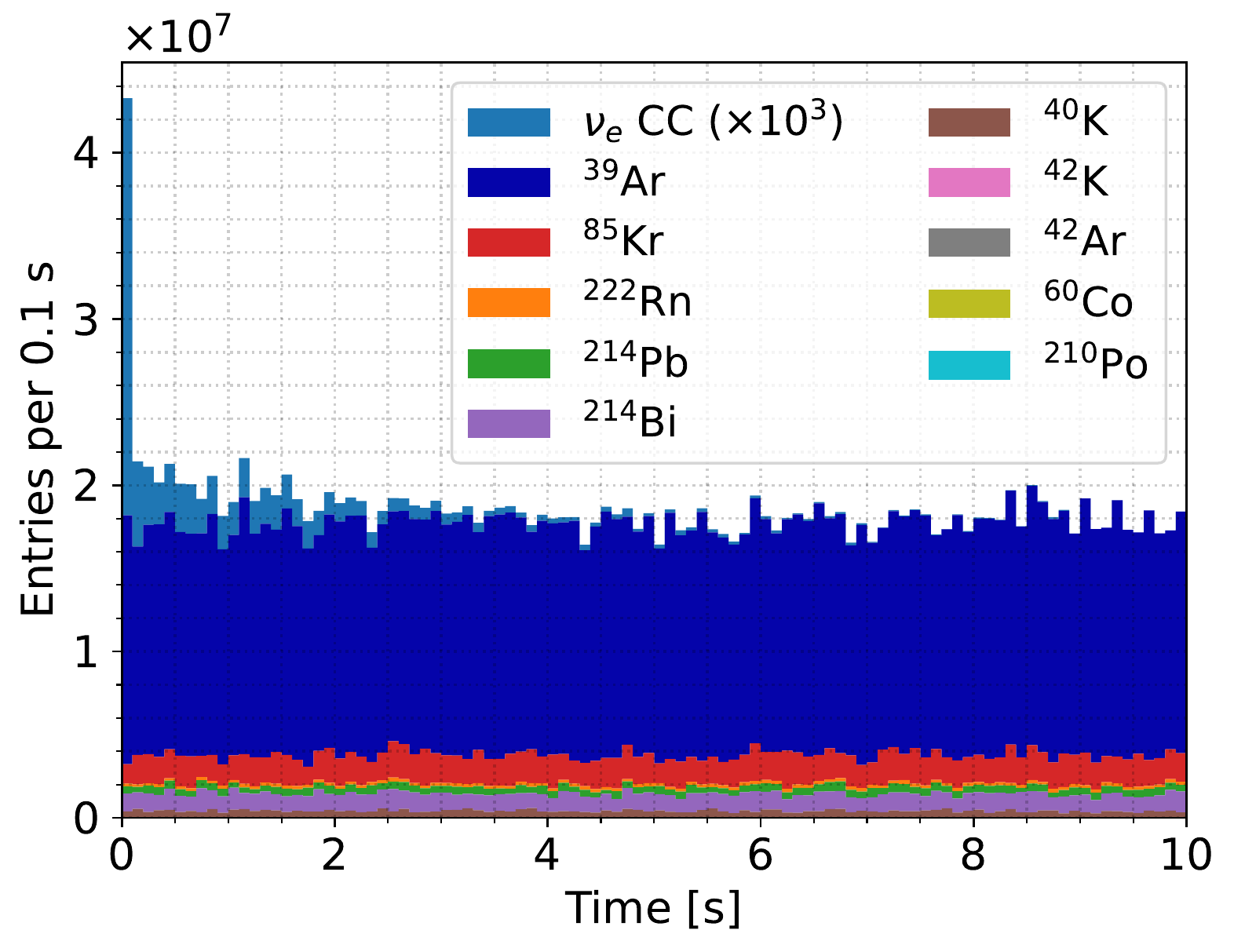}}}%
     \caption{Left: stacked histogram of the \textsc{Geant}4 truth-level deposited energy from $\nu_e$~CC signal events (magnified by $\times 10^3$) in a typical supernova (SN) simulation over 10 seconds and from the radiogenic background events.  Note that ionization charge quenching of $\alpha$ particles in LAr is not taken into account here, so the BiPo and RnPo signatures are higher than they should be. Right: stacked histogram of the timing profile of \textsc{Geant}4 hits from simulated SN signal and radiogenic background events.}
     \label{fig:supernovawithbackground}
 \end{figure*}

To realistically model the background events, the radioactive decays are simulated from the location of the material which produces them (see Table~\ref{tab:radiogenic-backgrounds} for the detailed locations). For example, the isotopes that decay in the LAr are placed uniformly and randomly in the liquid at the appropriate rate. When the source of the background is listed as coming from the anode-plane assembly or cathode-plane assembly, the background is generated originating from those planes with the geometry in accordance with the DUNE far detector technical design report (TDR)~\cite{DUNETDR}. In the case where an isotope decayed to a progeny that is unstable, the progeny is then allowed to decay with its characteristic lifetime. This is particularly important in the case of isotopes such as \ce{^{214}Bi} which decay to \ce{^{210}Po}, which itself has a short half-life. As noted in Section~\ref{sec:Q-PixOverview}, the envisioned geometry of a 10-kton Q-Pix module is slightly modified from the one presented in the DUNE FD TDR, specifically the APA is reduced in size and mass. However, the results presented here do not include this modification for simplicity of presentation and comparisons to other DUNE-related works. While the pixel-based readout does add more mass and potentially more sources of radiogenic backgrounds via the inclusion of printed circuit boards (PCBs), the discrimination power afforded by the Q-Pix readout (shown in Section \ref{sec:EventID}) in conjunction with the ability to produce low-background PCBs \cite{NEXT:2014giv,Akerib:2008zz,Iguaz:2015myh} provides confidence that this can be effectively mitigated. Fig.~\ref{fig:supernovawithbackground} shows the energy and timing spectra of the signal and backgrounds used in this analysis.

\subsection{Q-Pix Simulation}\label{sec:Q-PixSimulation}
The simulation framework developed for Q-Pix primarily consists of two {\tt C\texttt{++}}-based packages: (1) a \textsc{Geant}4-based~\cite{Agostinelli:2002hh, Allison:2006ve, Allison:2016lfl} package for simulating the interactions and ionization of particles within a liquid argon volume, and (2) a Q-Pix-specific package for converting \textsc{Geant}4 hits (a \textsc{Geant}4 \textit{hit} is defined here to be a segment of a simulated particle's trajectory and carries information about its position, timing, and energy deposited) into ionization electrons that are then propagated to a pixel plane where the pixel response is simulated.  The framework simulates single DUNE-SP APA drift volumes (one volume consists of a $2.3$~$\times$~$6$~m$^2$ charge collection/readout plane with a $3.6$~m drift length) which are then analyzed individually.  A 10-kton module's sensitivity can then be extracted by joining 200 APA drift volumes; performing the simulation in this manner was required to keep the files produced to a manageable size. 

The Q-Pix-specific ``Reset-Time Difference'' ({\tt RTD}) package simulates the response of the Q-Pix electronics. This is done by first converting the \textsc{Geant}4 hits into ionization electrons with the assumed LAr W-value of 23.6~eV/electron. Next, the simulation accounts for recombination using the ``modified box" model~\cite{ArgoNeuT:2013kpa}, and removes those electrons from consideration. The electrons that remain are then uniformly distributed between the start and end of the \textsc{Geant}4 hit. At this point the drift time for each electron is calculated. This drift time is used to account for a reduction in the arriving signal due to electron lifetime as well as to smear the position of the electrons according to the longitudinal and transverse diffusion coefficients. This process is done for every ionization electron coming from a \textsc{Geant}4 hit in an event. All the electrons in a hit are sorted by drift time. The electrons are then subdivided into groups based on their $(x,y)$ position corresponding to which 4~$\times$~4~mm$^2$ pixel they will land on. This results in an array of hit pixels, each containing an array of time-sorted electrons. 

\begin{figure*}[htb]
    \centering
    \includegraphics[width=\textwidth]{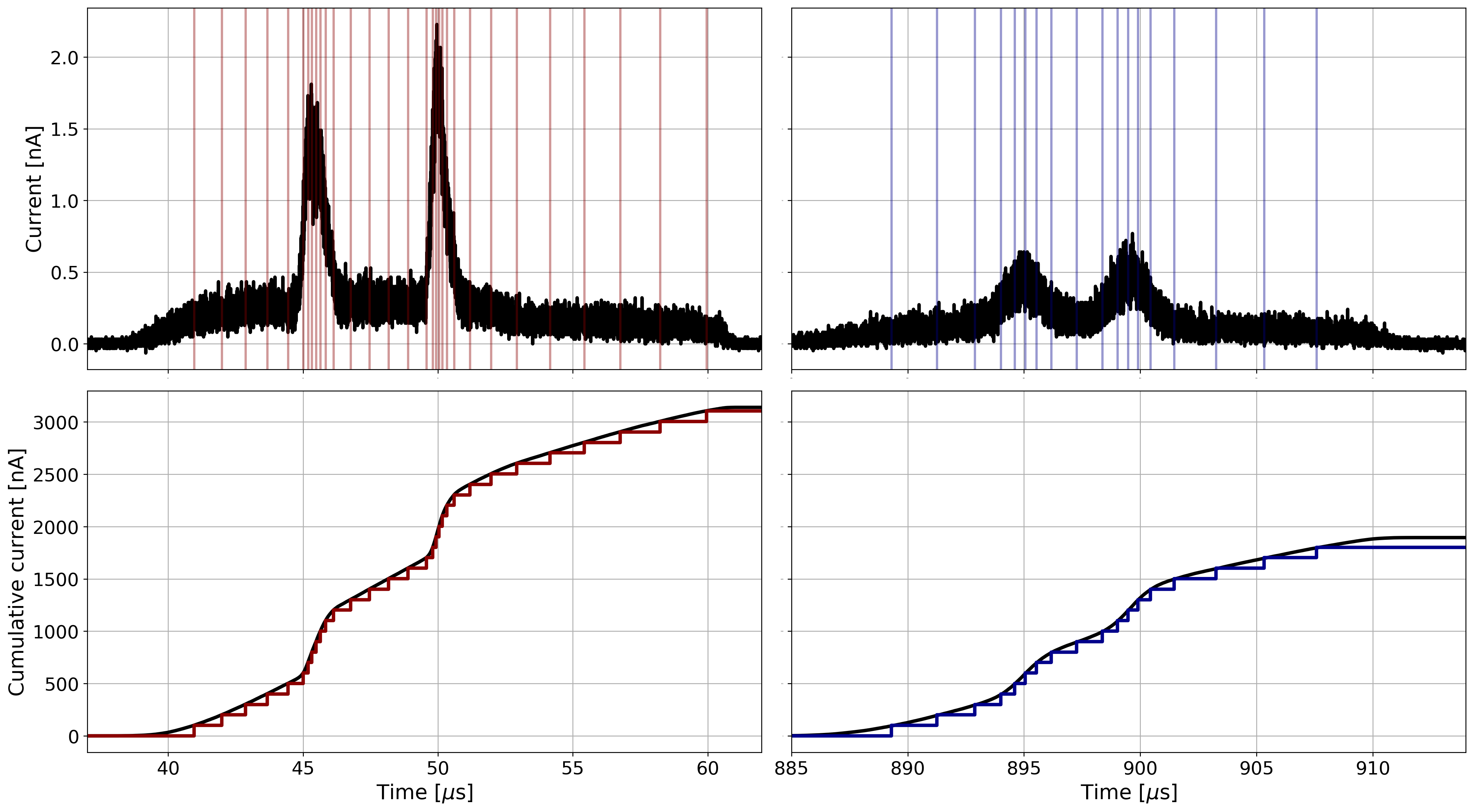}
    \caption{Examples of the current profiles (black lines) produced with the Q-Pix simulation package. The corresponding resets can be seen as the vertical lines and represent a 1~fC reset threshold. Left: current profile of the most active pixel of a 30~MeV electron event that is located 10~cm away from the pixel plane and launched perpendicular to the pixel plane. Right: current profile for the same event with the location of the electron 150~cm away from the pixel plane. One can see the expected broadening of the input signal due to the simulated diffusion as well as the reduction in the integrated current.}
    \label{fig:QPIX_RTD_sim}
\end{figure*}

Producing the associated Q-Pix resets is then done by integrating the charge on each pixel with a time step of 10~ns. The time step size is assumed to be exact for simplicity. This assumption is justified since any 10~ns deviation would be sub-leading to a measurement of which is microseconds in length.  At each time step, the equivalent noise charge is added to mimic the current ASIC simulation of the Q-Pix front-end. The leakage current present on the front-end is a sub-leading noise component. Without a prototype chip produced, the actual value of the leakage current is unknown. For the purposes of this study, we assume it to be 100~aA. This number is comparable to the leakage current measured by the LArPix ASIC~\cite{Dwyer:2018phu}, which shares similar technology and geometric considerations to Q-Pix. Finally, in the absence of a prototype to benchmark, we further assume no charge is lost or perturbed during a reset. Such an assumption can be realized in the ASIC using a various techniques under exploration (e.g., charge-pump front-end). Ultimately, the impact of any charge loss can be calibrated and is thus taken as a safe assumption. When a pixel collects enough electrons to undergo a reset (e.g., 6250~electrons, or 1~fC), the time and pixel number is logged and the electrons are drained from the pixel. Performing the simulation in this manner closely resembles the actual electronics response as well as the data format that is produced. The pixel ID and reset times are then stored in the same {\tt ROOT}~\cite{Brun:1997pa} file as the \textsc{Geant}4 information. The different parameter values used in the simulations described above are listed in Table~\ref{tab:simulated_paramaters}. The {\tt RTD} code can be configured to produce current profiles, which can be exported and use as inputs to other simulation software. An example current profile with corresponding reset stamps can be seen in Fig.~\ref{fig:QPIX_RTD_sim}, which illustrates the same event placed at two drift distances (e.g., 10~cm and 150~cm). The change in the current profile, and thus the subsequent number and frequency of the resets, can be seen due to this change in drift distance.

Finally, we note here that the ionization charge produced from $\alpha$ particles in a liquid noble element medium tends to be relatively small and more scintillation light is produced~\cite{Mei:2007jn}.  This ionization charge quenching is not taken into account in the Q-Pix simulation, so the BiPo and RnPo signatures in the energy spectrum on the left in Fig.~\ref{fig:supernovawithbackground} are higher than they should be.  The Noble Element Simulation Technique (NEST)~\cite{Szydagis:2011tk, Szydagis:2013sih, Lenardo:2014cva} software package, which is capable of modeling the ionization charge quenching in LAr, will be integrated into the Q-Pix simulation for future work.

\begin{table}[ht]
    \smaller
    \centering
    \setlength{\tabcolsep}{10pt}%
    \def\arraystretch{1.2}%
    \resizebox{0.89\textwidth}{!}{%
    \begin{tabular}{lc}
        \toprule
        Parameter    & Value \\
        \midrule
        W-value for ionization in LAr    & 23.6~eV/e$^-$     \\
        Drift electric field             & 500~V/cm          \\
        Drift velocity                   & 164,800~cm/s      \\
        Longitudinal diffusion           & 6.8223~cm$^2$/s   \\
        Transverse diffusion             & 13.1586~cm$^2$/s  \\
        Electron lifetime                & 0.1~s             \\
        \multirow{2}{*}{Reset threshold} & 1~fC~(6250~e$^-$, or \\
                                         & 147.5~keV in LAr) \\
        Sampling time                    & 10~ns             \\
        \bottomrule
    \end{tabular}
    }
    \caption{Summary of the physical parameters used in the Q-Pix simulations.}
    \label{tab:simulated_paramaters}
\end{table}

\section{Results}\label{sec:Results}
In this section, we present the results of the supernova simulation done with the Q-Pix readout and show how this readout architecture can enhance supernova burst reconstruction for large-scale LArTPCs. In Section~\ref{sec:EnergyReconstruction}, we give an overview of the energy reconstruction and, in particular, the improvement seen in low-energy events. Section~\ref{sec:EventID} provides the algorithm developed to identify supernova events and distinguish them from background. This identification is not the same as a trigger efficiency since, as highlighted in Section~\ref{sec:DataRates}, the amount of data to readout an entire 10-kton module is remarkably low. Section~\ref{sec:Triggering} provides details for taking the identified events and understand what additional criteria needs to be applied to allow for a large-scale LArTPC using Q-Pix to serve as a trigger to the SuperNova Early Warning System (SNEWS)~\cite{Antonioli:2004zb} and the associated supernova burst trigger efficiency. Finally, Section~\ref{sec:Direction} discusses the use of the intrinsic 3D information provided by Q-Pix to reconstruct the direction of the supernova source from an analysis of the neutrinos detected.

\subsection{Energy Reconstruction}\label{sec:EnergyReconstruction}

In order to compare this work to others, an energy reconstruction conversion factor is derived. The energy is reconstructed from collected charge and compared to true neutrino energy. In this conversion factor, the reconstructed energy fraction depends on 3D information of the event, drift, diffusion, and topology. From this conversion factor, a correction matrix based on the 3D reconstruction of the charge is generated, similarly to the method used in Ref.~\cite{DUNE:2020zfm}. The correction matrix takes into account the generation, transport, and detection of ionization signals in energy steps of 0.1~MeV and drift steps of 50~cm. This was done to get a fine energy sampling at low energy ($<$5~MeV) and to account for the drift and diffusion of such events throughout the detection volume. Single electrons were generated isotropically at the various energies relevant to supernova and drift distances. From this dataset, parameterizations are computed to produce the correction matrix of the true energy for a given number of resets observed as a function of drift distance. In this work, the unspecified Q-Pix photon detection system is assumed to provide the drift distance with an uncertainty of $\pm 25$~cm. This correction only applies to this section for comparing to other works by correcting for drift-dependent attenuation of the deposited energy relative to the true energy.

\begin{figure*}[htb]
    \centering
    \includegraphics[width=0.99\textwidth]{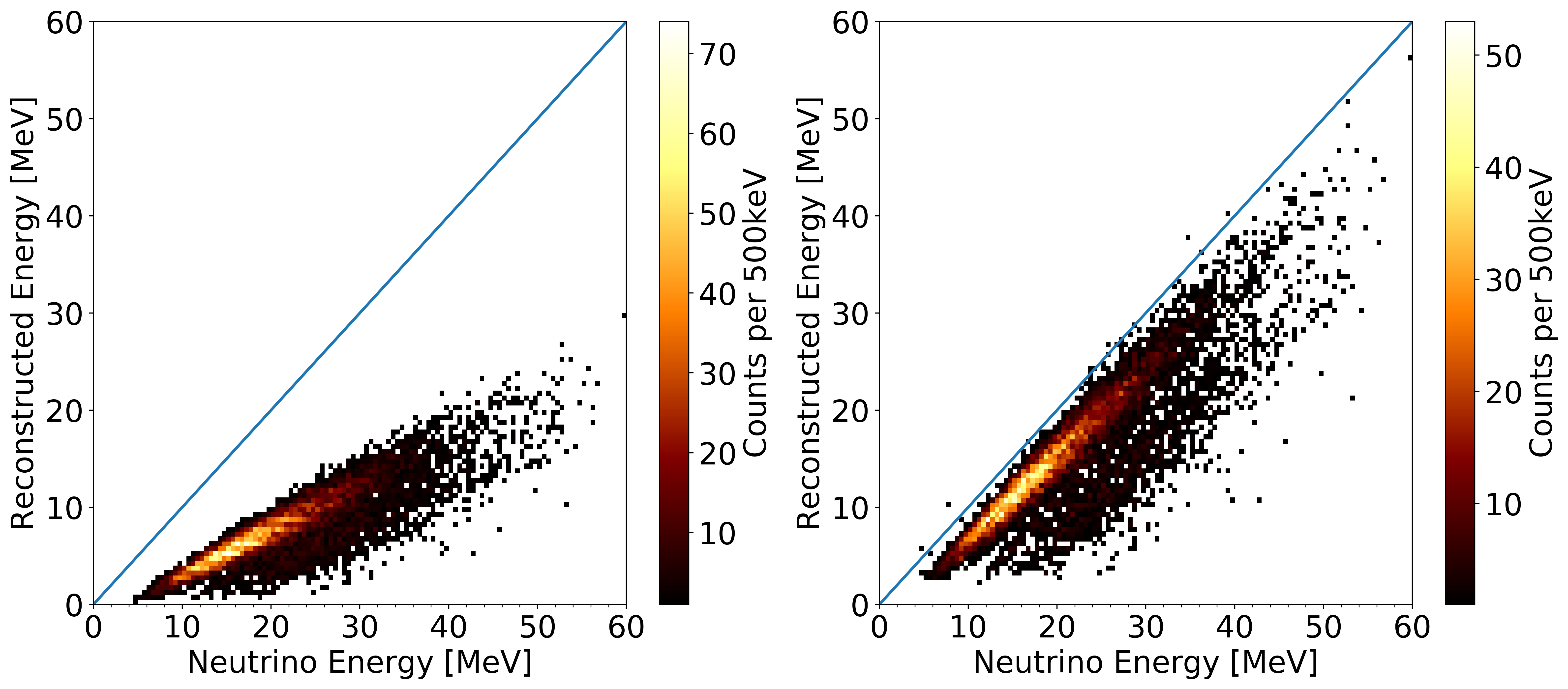}
    \caption{Left: the reconstructed energy for supernova events as a function of true neutrino energy with no energy correction applied. Right: the reconstructed energy for supernova events as a function of true neutrino energy after application of the energy correction matrix which takes into account electron lifetime, diffusion, event topology, and the mapping between deposited energy and true energy as a function of drift distance.}
    \label{fig:EnergyDepositedvsNeutrinoEnergy}
\end{figure*}

Fig.~\ref{fig:EnergyDepositedvsNeutrinoEnergy} shows the reconstructed energy as a function of the true neutrino energy simulated with {\tt MARLEY} with and without the energy corrections. While the uncorrected energy distribution is linear as a function of true neutrino energy, there is a clear offset and difference in slope. By the application of the correction matrix the slope is increased, while an overall offset remains.

\begin{figure}[htb]
    \centering
    \includegraphics[scale=0.15]{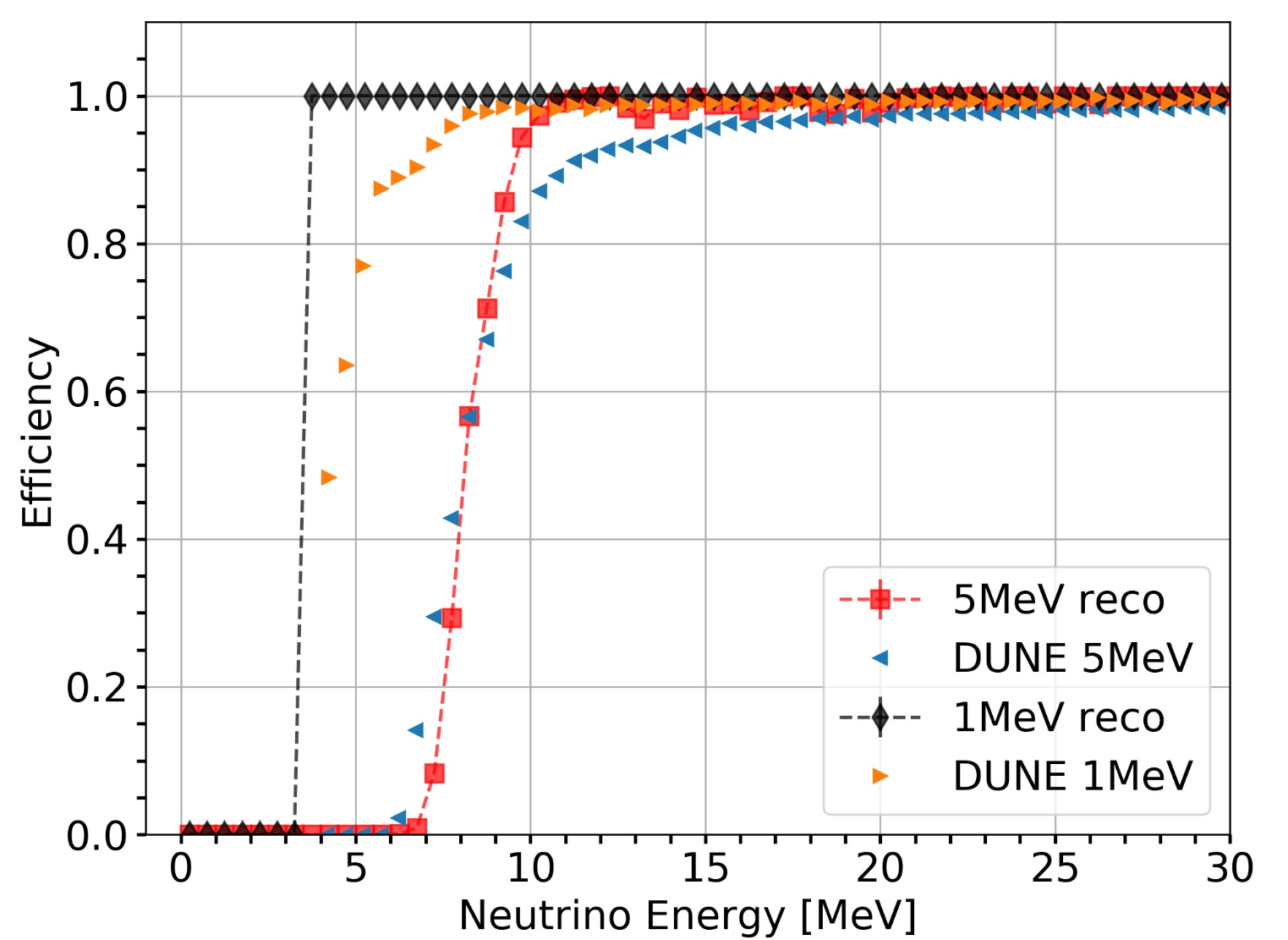}
    \caption{Supernova event reconstruction efficiency as a function of true neutrino energy for different minimum total deposited energy requirements, or energy thresholds. For comparison, the efficiencies for DUNE from Ref.~\cite{DUNE:2020zfm} are shown in blue triangles for a minimum of 5~MeV deposited and in orange triangles for a minimum of 1~MeV deposited.}
    \label{fig:LowEnergyRecoEnhancement}
\end{figure}

Fig.~\ref{fig:LowEnergyRecoEnhancement} shows the event reconstruction efficiency for two different energy thresholds. 
The Q-Pix architecture increases the reconstruction efficiency at lower neutrino energy compared to the results presented for DUNE in Ref.~\cite{DUNE:2020zfm}. The efficiency rises to nearly 100\% very rapidly and maintains this high efficiency down to lower neutrino energy. Given the current level of simulation, it appears that a Q-Pix-based readout can successfully identify and reconstruct very low energy supernova neutrino interactions, and has a performance in this task comparable to that exhibited in existing DUNE wire readout simulation and reconstruction.

\subsection{Event Identification}\label{sec:EventID}
\begin{figure*}[h!]
    \centering
    \includegraphics[width=0.95\textwidth]{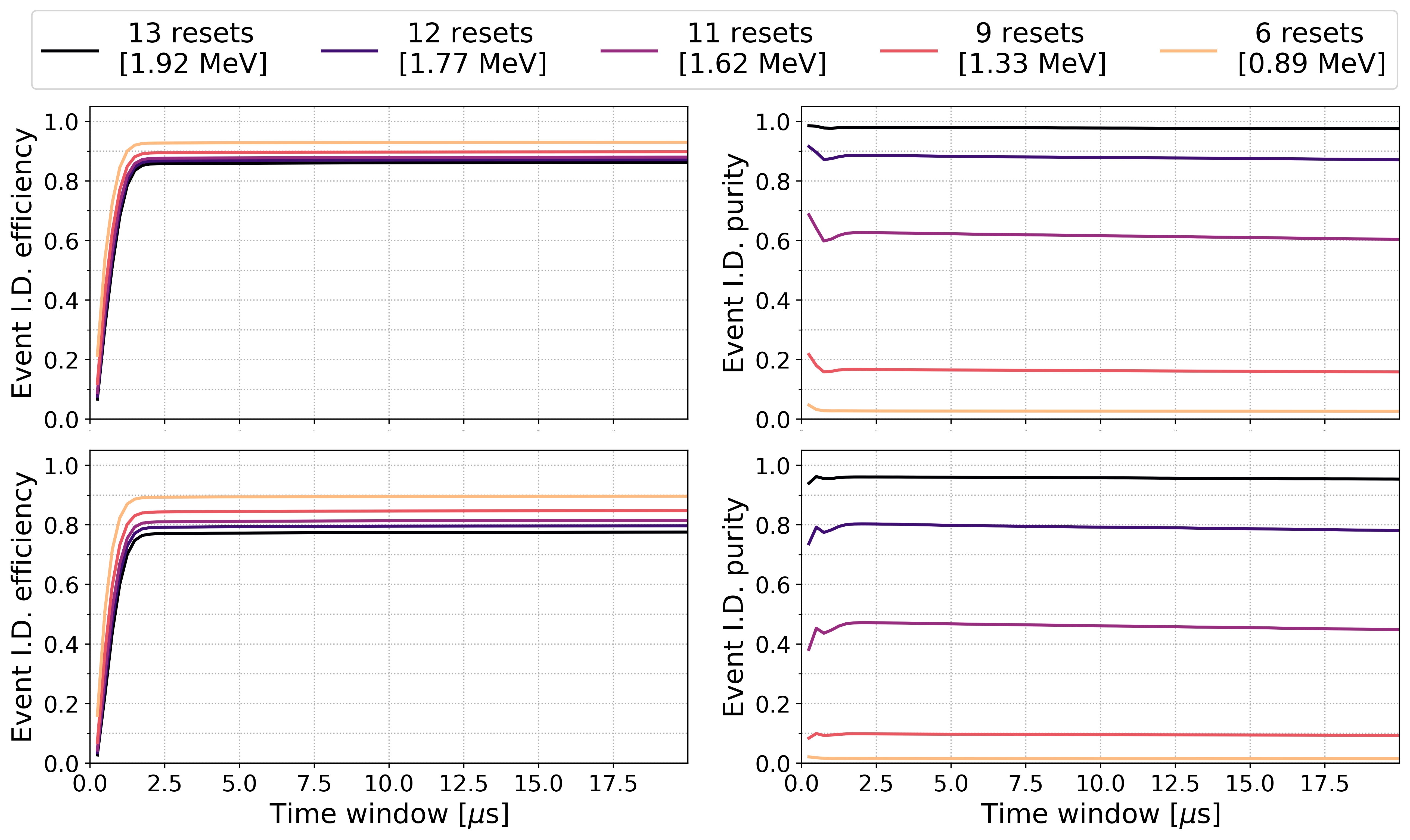}
    \caption{Top row: event ID efficiency (left) and purity (right) for the full supernova energy spectrum. The colors represent different reset thresholds. Bottom row: same as above, but only considering events with less than 5~MeV of energy in order to optimize the background reduction. Both are plotted as a function of the cluster time window.}
    \label{fig:efficiency/purity/time}
\end{figure*}

\begin{figure*}[h!]
    \centering
    \includegraphics[width=0.95\textwidth]{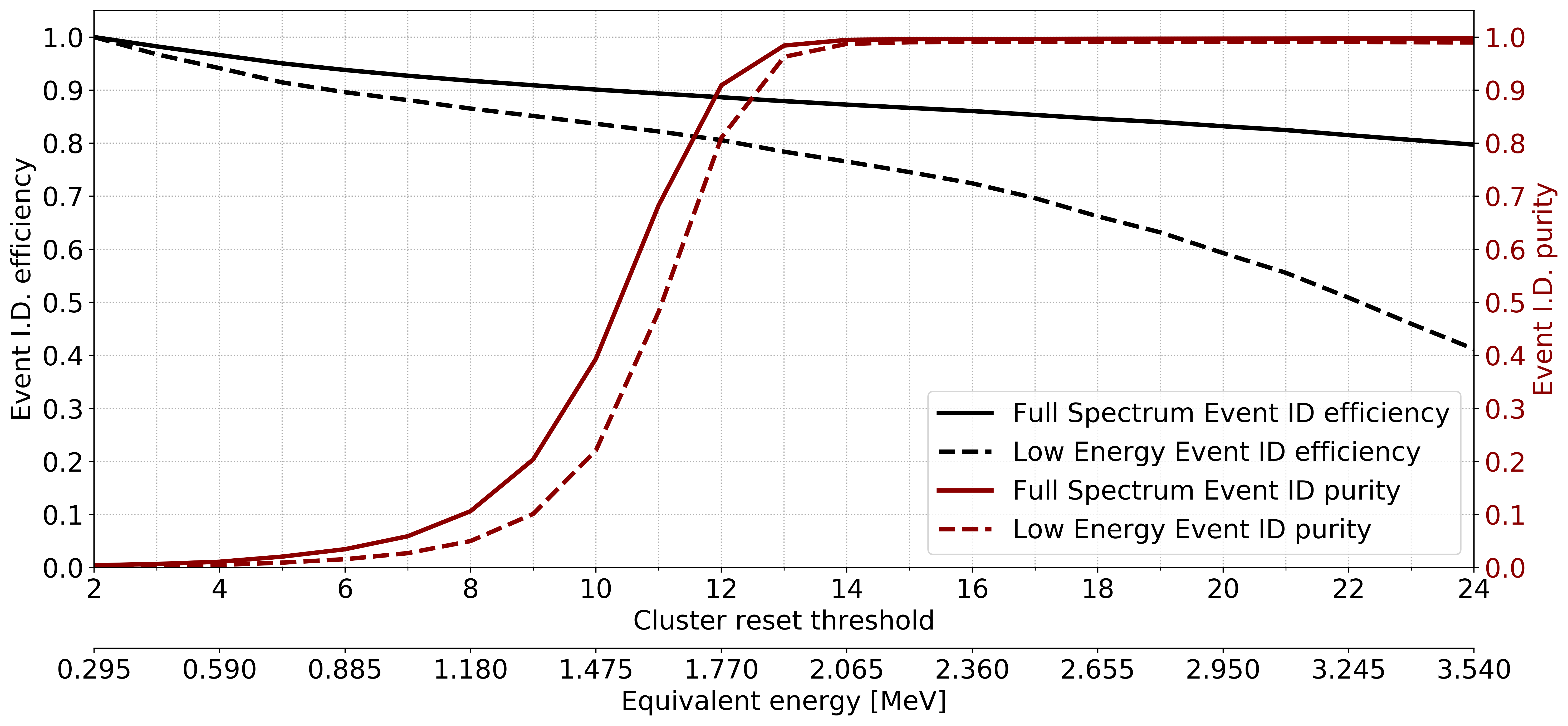}
    \caption{Event ID efficiency (black) and purity (red) as a function of minimum number of resets per cluster for the 3$\mu$s time window. Results are shown for the full energy spectrum (solid line) and for the low-energy events ($\leq$5~MeV) (dashed line).}
    \label{fig:Event_ID_reset}
\end{figure*}

Due to the addition of radiogenic backgrounds in this work, it is necessary to define an identification scheme to separate the radiogenic backgrounds from the supernova signal events. This is done by taking advantage of the inherent 3D information of the pixelization, as well as the unique timing information that Q-Pix produces. This allows for a rather simple algorithm that clusters the resets together to build 3D trajectories (referred to as tracks). By selecting on the number of resets required to form a 3D track, the radiogenic backgrounds can be distinguished from the supernova neutrino interactions due to the track extent and energy. 

However, there is an optimization on where to place this selection in time between resets, number of resets in a track, and how localized these resets are in space. For spatial localization, a simple assumption is made that only neighboring 4~$\times$~4~mm$^2$ pixels are considered when grouping together pixels. This means that if a given set of energy depositions has a sufficient distance in space and time it will be clustered into two distinct tracks. An example of this in a supernova interaction is energy deposition due to de-excitation photons that can be further away than the primary electron track and are likely to produce separate tracks. This ultimately leads to a limit in this analysis achievable efficiency as the focus of the reconstruction is to distinguish between radiogenics and deposits due to supernova neutrinos. 

To optimize the remaining two parameters (i.e., the time between resets and number of resets in a track) a sample of 10,000 unique background sets of events along with a sample 10,000 unique supernova interactions were generated in a single APA volume. It is important to note here that these two samples are not mixed (i.e., the events analyzed are either purely background or purely signal), and the optimization is performed on an isolated background or signal event for simplicity. As clearly seen on the left in Fig.~\ref{fig:supernovawithbackground}, radiogenic background events are problematic mostly for events with $<$5~MeV of deposited energy. We therefore focus on the optimization of the parameters with low-energy events (e.g., total deposited energy $\leq$5~MeV).

The event identification efficiency is calculated as the number of reconstructed signal resets in a given time window divided by the total number of true signal resets in an event, where the number of signal resets in a time window is the summed number of resets of all tracks which contain more resets than the chosen threshold. In a similar manner, the event identification purity is defined as the number of reconstructed signal resets in a time window divided by the total number of reconstructed signal and background resets. An example of such efficiency and purity can be seen in Fig.~\ref{fig:efficiency/purity/time}, where each line represents a different reset thresholds for a given track. From Fig.~\ref{fig:efficiency/purity/time}, the optimal time window between any two resets is chosen to be 3~$\mu$s. While a slightly shorter or significantly longer time could be chosen without impacting the efficiency or purity, 3~$\mu$s is the $3\sigma$ longitudinal diffusion threshold for a point source diffusing in LAr from 3.6~m away in a drift electric field of 500~V/cm. The plateau in efficiency shown in Fig.~\ref{fig:efficiency/purity/time} at $\sim90\%$ results from the clustering algorithm ``missing'' charge from de-excitation photons which radiate further away than the 4~mm pixel pitch. More advanced spatial clustering algorithms could improve this, but this choice was found sufficient to accept the supernova neutrino signals while continuing to reject radiogenic backgrounds.

After choosing the optimal 3~$\mu$s time window between resets, we next define the threshold on the number of resets per cluster. Using the time window of 3~$\mu$s, the event identification efficiency and purity as a function of the threshold is shown in Fig.~\ref{fig:Event_ID_reset}. For the events with $\leq$5~MeV of total deposited energy a purity of $>$95\% can be achieved at an efficiency of $\sim$80\% with a threshold of 13 resets (equivalent to $\sim$1.85~MeV deposited energy). The purity across the full supernova neutrino energy spectrum is $>$99\%  for an efficiency of $\sim$88\% when using the 13 reset threshold. This last point of optimization can be tuned based on the analysis for different purity and efficiency choices.

\subsection{Data Rates}\label{sec:DataRates}
A challenge for supernovae detection with a kiloton-scale LArTPC is managing data rates for low-energy thresholds of $<$10~MeV. One of the remarkable outcomes of the Q-Pix principle of electronic least action is the ability of the detector to achieve low-energy thresholds while maintaining very low data rates.

\begin{figure}[htb]
    \centering
    \includegraphics[width=\textwidth]{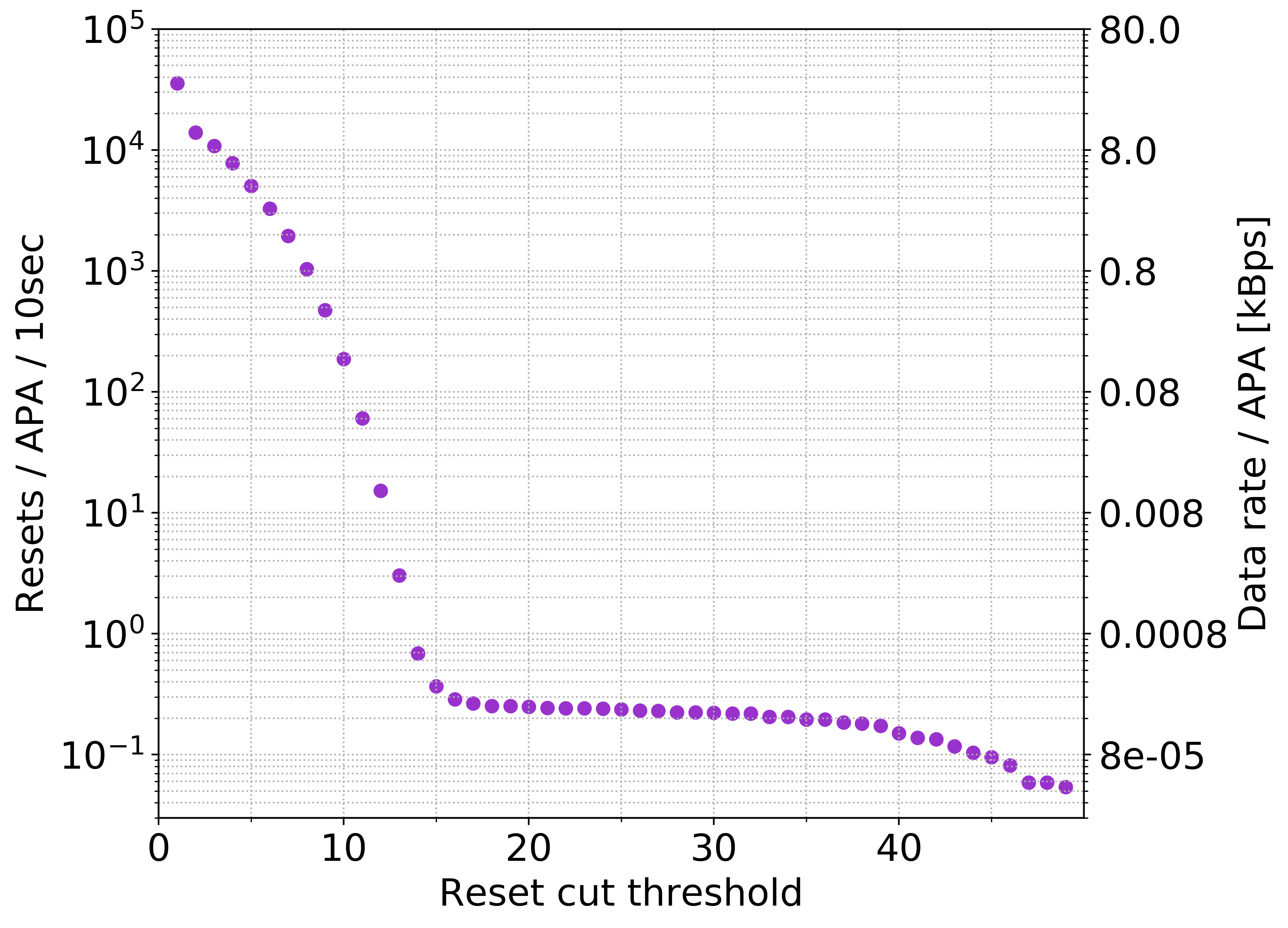}
    \caption{The data rates shown here are the average of 10,000 unique 10-second-long APA drift volumes as a function of the threshold on the number of resets. Data rates for one APA from radiogenic background events will be a function of the threshold on the number of resets. The rates are shown in number of resets per APA per 10 seconds (left axis) and the corresponding data rate (right axis).} 
    \label{fig:QPIX_DataRate}
\end{figure}

Fig.~\ref{fig:QPIX_DataRate} shows the data rate per APA based on the radiogenic backgrounds outlined in Table~\ref{tab:radiogenic-backgrounds}. The rate is calculated from the average number of resets observed in an APA over a 10-second readout window using 10,000 unique sets of radiogenic backgrounds. On the left axis of Fig.~\ref{fig:QPIX_DataRate}, the average data rate is calculated per second per APA assuming each reset is encoded by 64~bits of information. Even at the lowest threshold of 1~reset (which corresponds to 147~keV of deposited energy) the data rate per full 10-kton module (assuming 200 APAs for the Q-Pix readout) is only 5.7~MB/s. This number drops by two orders of magnitude if 7~resets ($\sim$1~MeV of deposited energy) is required. Moreover, since the rate of radiogenics dominates, the inclusion of the supernova burst events from the previous sections leaves this estimated data rate largely unchanged.

In order to compare these data rates to the ones from the current DUNE FD design, some context is needed. First, the envisioned Q-Pix 10-kton module consisting of 200 APAs with 4~mm pixel pitch means there are $\sim$172 million channels to be compared to the 384,000 channels for the projective wire readout for the DUNE 10-kton module. This has to be taken into account when comparing data rate per channel. Second, the TPC data rates in the DUNE FD TDR~\cite{DUNETDR} assume 100 seconds of readout, a threshold of 10~MeV neutrino energy ($\sim$5~MeV of deposited energy), and are quoted as an annual data volume. As a comparison, the numbers calculated above for Q-Pix in this paper assume much lower energy thresholds and are calculated per second. With this context, Table~\ref{tab:dataratecompare} presents data rates for the Q-Pix architecture that can be directly compared to the ones predicted for the DUNE projective readout. 

\begin{table}[ht!]
    \setlength{\tabcolsep}{5pt}%
    \centering
    \resizebox{\textwidth}{!}{
    \begin{tabular}{c|c|c}
      \textbf{System}   & \makecell{\textbf{Data rate per 10 kton} \\ \textbf{per year (petabytes)}} & \makecell{\textbf{Data rate per channel} \\ \textbf{per second (kilobytes)}} \\
      \hline
      \makecell{Q-Pix 10 kton \\ pixel readout} & $1.03 \times 10^{-6}$ & $1.9 \times 10^{-10}$ \\
      \hline 
      \makecell{DUNE 10 kton \\ projective readout} & $<$2 & 1.6 \\
    \end{tabular}}
    \caption{Comparison of data rates between a 10-kton DUNE projective readout as described in Ref.~\cite{DUNETDR} and a Q-Pix 10-kton module described in this work. In both cases, a 10~MeV energy threshold is assumed.}
    \label{tab:dataratecompare}
\end{table}

For the Q-Pix readout, assuming a threshold of 10~MeV neutrino energy ($\sim$5~MeV of total deposited energy, or 34~resets), the data rate from the dominant source of radiogenic background events is estimated to be ${16.5\times 10^{-5}}$~kB/s/APA or equivalently 0.032~kB/s/10~kton which is 1.03~GB/year/10~kton. This last number is to be compared to the estimated DUNE FD data rates of less than 2~PB/year/10~kton, which is about six orders of magnitude larger.

\begin{figure*}[ht!]
    \centering
    \includegraphics[width=0.9\textwidth]{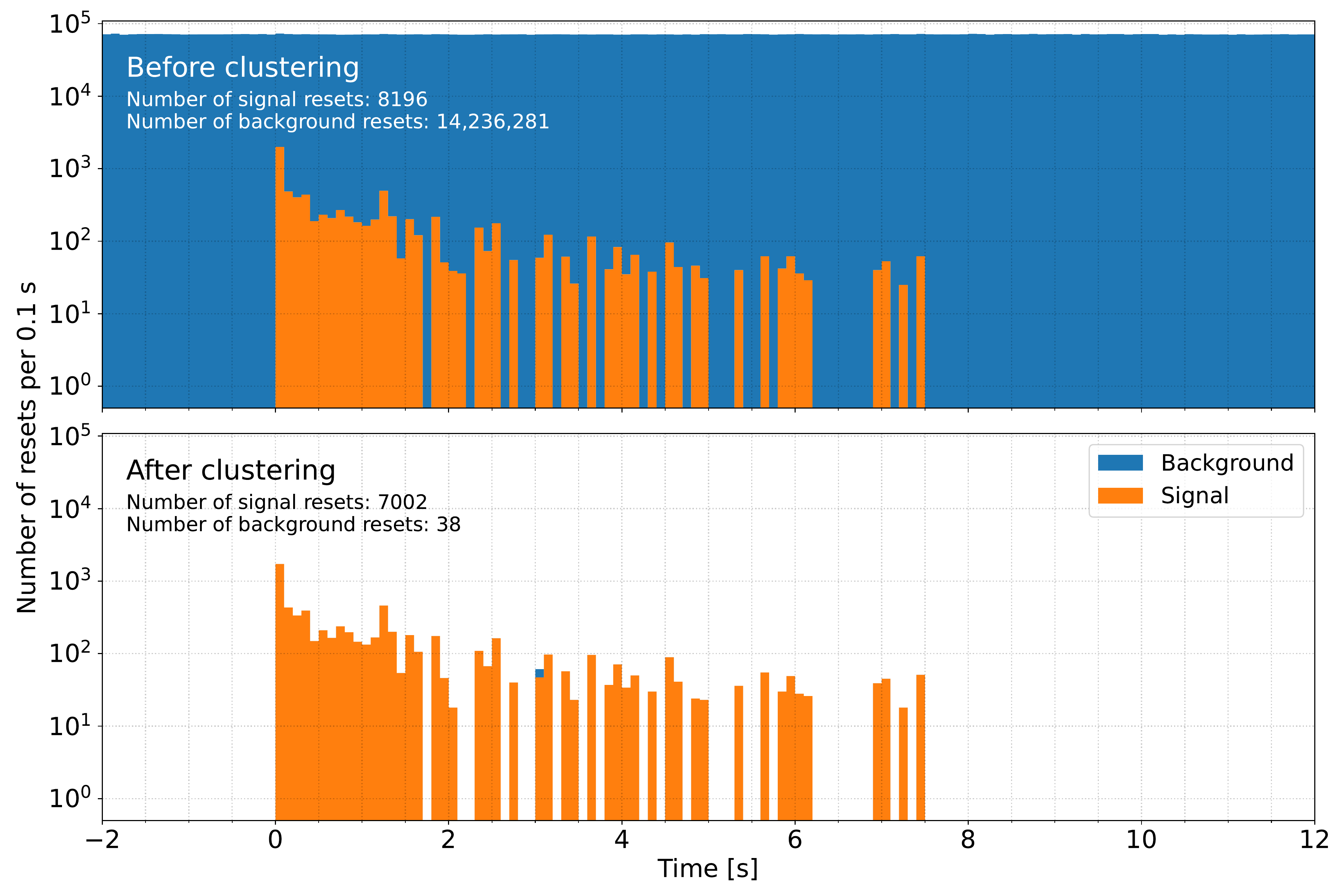}
    \caption{Reset times in a 10-kton Q-Pix detector module before (top) and after (bottom) clustering for a supernova burst (SNB) simulated at 10~kpc.  In this example, the SNB reset clustering efficiency is ${\frac{7002}{8196} \approx 85\%}$ and purity is ${\frac{7002}{7002 + 38} \approx 99\%}$.}
    \label{fig:snb-timing-cluster}
\end{figure*}

The anticipated full data rate from a DUNE projective readout 10-kton module with a 10~MeV neutrino energy threshold including beam events, astrophysical sources of neutrinos, and calibrations has been presented in various Refs.~\cite{Abi:2018dnh,Schellman:2020vdz,DUNETDR}, and is estimated to be between 25-30 petabytes (PB) per year. While beyond the immediate scope of this paper, it is very clear that the Q-Pix architecture can vastly reduce these data rates across all the different neutrino sources and can allow for much lower energy thresholds for comparable data rates (e.g., for Q-Pix, the data rate would be 0.18~PB/10~kton/year for a 147~keV threshold based on radiogenic backgrounds).

Furthermore, these anticipated low data rates coupled with the ability to store the resets locally and to have all the ASIC chips periodically read out could remove the need for a dedicated ``trigger'' whereby an external system forces the readout of all detector channels. However, in the case of punctual events, such as a supernova burst, it is of the utmost importance that the data can alert or trigger the scientific community. In the next section, we explore what such a ``trigger'' would look like within the Q-Pix architecture.

\subsection{Supernova Neutrino Burst Triggering}\label{sec:Triggering}

In the event of a supernova burst (SNB), the Q-Pix-enabled detector module must be sensitive to the increase in MeV-scale neutrino activity in the LAr detector volume and be capable of identifying that it is indeed a SNB event with high confidence.  This, paired with the ability to determine the direction of the supernova burst (discussed in Section~\ref{sec:Direction}), will allow a detector module with Q-Pix readout to be a contributor to the SuperNova Early Warning System (SNEWS). In this section, we define a ``trigger'' to be an alert given by a near real-time analysis of a possible supernova burst.

The identification of a supernova burst event in a Q-Pix-enabled detector module can be done by first clustering resets to achieve a high-purity sample of neutrino interactions, taking the sum of the number of resets within some time window, and then placing a minimum number of resets required for a ``trigger.'' 
The clustering algorithm used is based on DBSCAN~\cite{Ester96adensity-based} with a Chebyshev metric~\cite{10.5555/357286}; this algorithm can process up to five hundred 10-second time windows per minute which is more than sufficient to provide a near real-time identification of a supernova burst with a Q-Pix-enabled detector module.  Here, we run the clustering algorithm over simulated samples with a mix of both signal and background events so these parameters are optimized differently from Section~\ref{sec:EventID} (where the clustering is performed on individual neutrino interactions or radioactive decays) and thus more realistic.  The maximum $\Delta t$ between each reset (for a reset to be considered to be in the neighborhood of another reset) is optimized to 6~$\mu$s, and the minimum number of resets to form a cluster is optimized to 14 ($\sim$2.065~MeV).  An example of reset times in a single 10-kton Q-Pix-enabled detector module before and after clustering within a time window is shown in Fig.~\ref{fig:snb-timing-cluster} for a supernova burst at 10~kpc; the effectiveness of reducing the radiogenic background (shown in blue) from the supernova interactions (shown in orange) with very little contamination remaining can be quantified with a SNB reset clustering efficiency of $\approx$85\% and purity of $\approx$99\%.

\begin{figure}
    \centering
    \includegraphics[width=\textwidth]{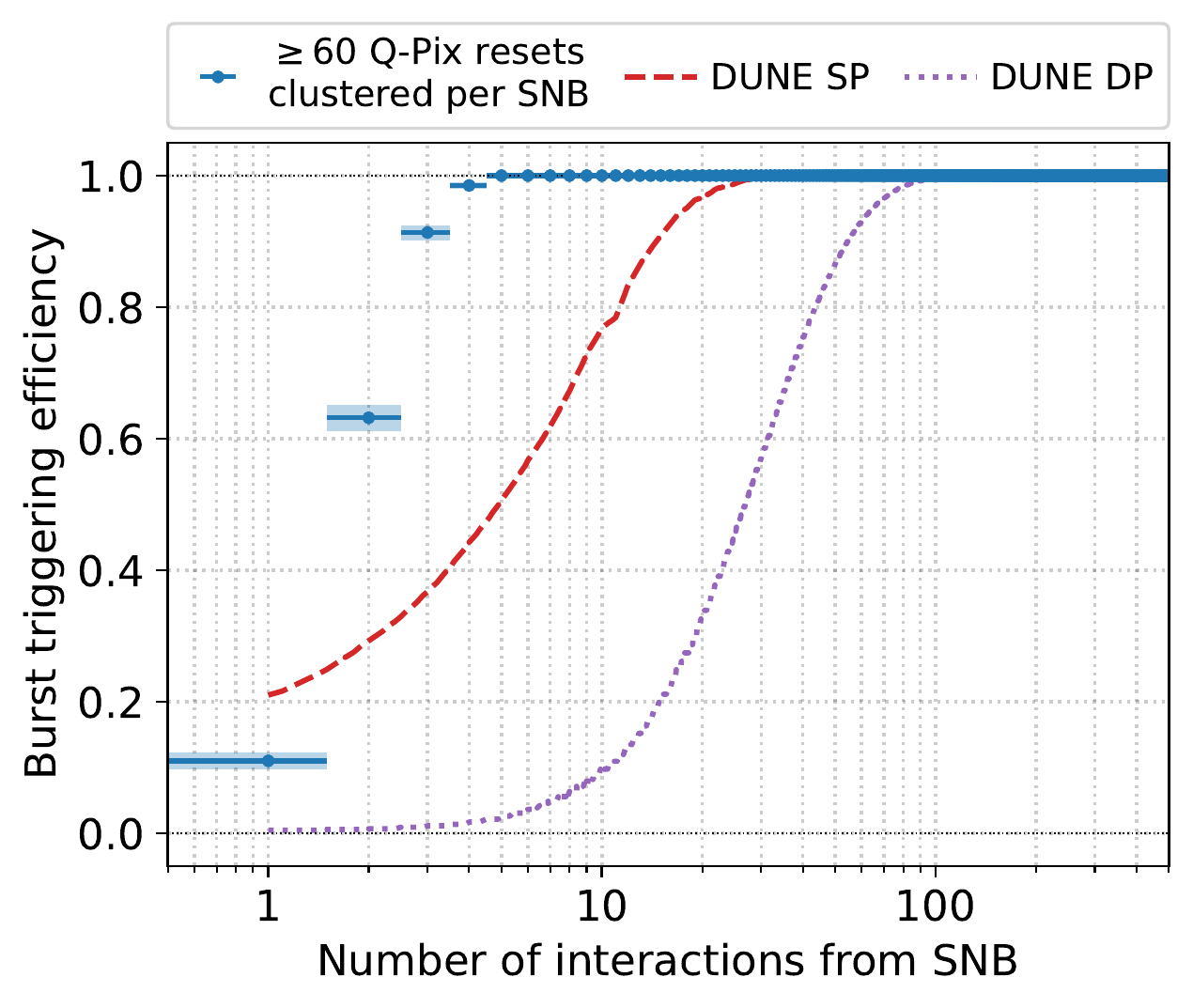}
    \caption{Supernova burst triggering efficiency as a function of the number of $\nu_e$ interactions in a 10-kton Q-Pix-enabled detector module.  The blue points indicate the case where the total number of Q-Pix resets after clustering within a 10-second window is required to be at least 60 ($\sim$8.85~MeV).  The DUNE SP (single-phase) and DP (dual-phase) efficiencies are shown for comparison and are taken from Ref.~\cite{DUNE:2020zfm}.}
    \label{fig:burst-triggering-efficiency} 
\end{figure}

The burst triggering efficiency, defined as the number of detected SNBs divided by the total number of SNBs, as a function of the number of neutrino interactions from a supernova burst is shown in Fig.~\ref{fig:burst-triggering-efficiency}.  A minimum of 60~resets required within a 10-second time window corresponds to a minimum visible energy of $\sim$8.85~MeV and yields less than one fake trigger per month.  This triggering scheme allows a Q-Pix-enabled detector module to be a more efficient supernova burst detection module than the baseline DUNE single-phase (SP) or dual-phase (DP) detectors and currently envisioned triggering schemes, which are described in detail in Ref.~\cite{DUNE:2020zfm}.  This enhancement in efficiency means that with fewer supernova burst events, a Q-Pix-enabled detector module can still faithfully trigger on the presence of the excess activity caused by a supernova burst.  This expands the distance at which a supernova burst could occur for the detector to still be capable of identifying it.  This highly efficient trigger can easily be achieved for a supernova occurring at distances that would yield tens of neutrino interactions ($\mathcal{O}(60\textrm{-}70)$~kpc) in a single 10-kton module.

\subsection{Directionality Determination}\label{sec:Direction}
The intrinsically high spatial resolution of LArTPCs, both wire-based and pixel-based, has the ability to determine the 3D spatial topology of the events. This enables the possibility of reconstructing the direction from which neutrinos originate, and in this section, we demonstrate for the first time how such a pointing analysis can be done with a LArTPC, specifically with the Q-Pix readout.


The ability to perform ``neutrino pointing'' that could enable multi-messenger astronomy is accomplished by providing directional information about where in the sky astronomers should look~\cite{Linzer:2019swe, Beacom:1998fj,PhysRevD.45.3361}. The neutrino signal from a collapsing star emerges on very prompt timescales [$\mathcal{O}$(seconds)] while the electromagnetic signal emerges on much slower timescales [$\mathcal{O}$(hours-days)]. Thus directional information can be used to provide an ``early warning''~\cite{Antonioli:2004zb} to astronomers. Moreover, as it has been pointed out in Refs.~\cite{OConnor:2010moj,Keehn:2010pn,Mukhopadhyay:2020ubs}, some supernovae may not produce any obvious electromagnetic signature while still producing copious amounts of neutrinos. These so-called ``failed'' supernovae could still be identified if the field of search could be narrowed by neutrino pointing. Finally, even without identifying the astronomical source of the neutrinos, directionality determination of the neutrinos can help evaluate the trajectory the neutrinos took en route to detection and thus allow for estimates of neutrino matter effects originating from their interaction with the Earth~\cite{Tomas:2003xn}.

\begin{figure*}[htb!]
    \centering
    \includegraphics[width=0.48\textwidth]{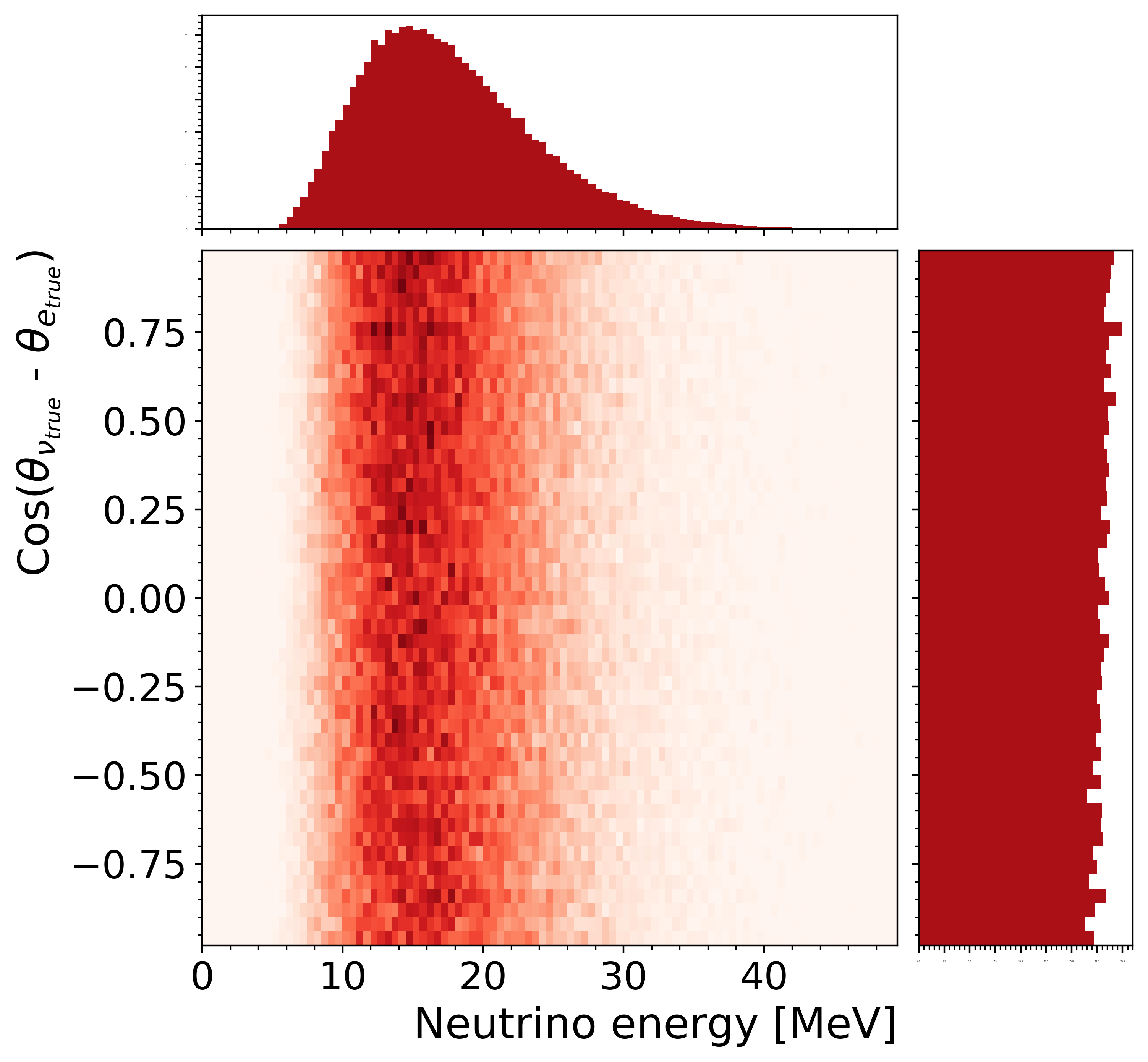}
    \includegraphics[width=0.48\textwidth]{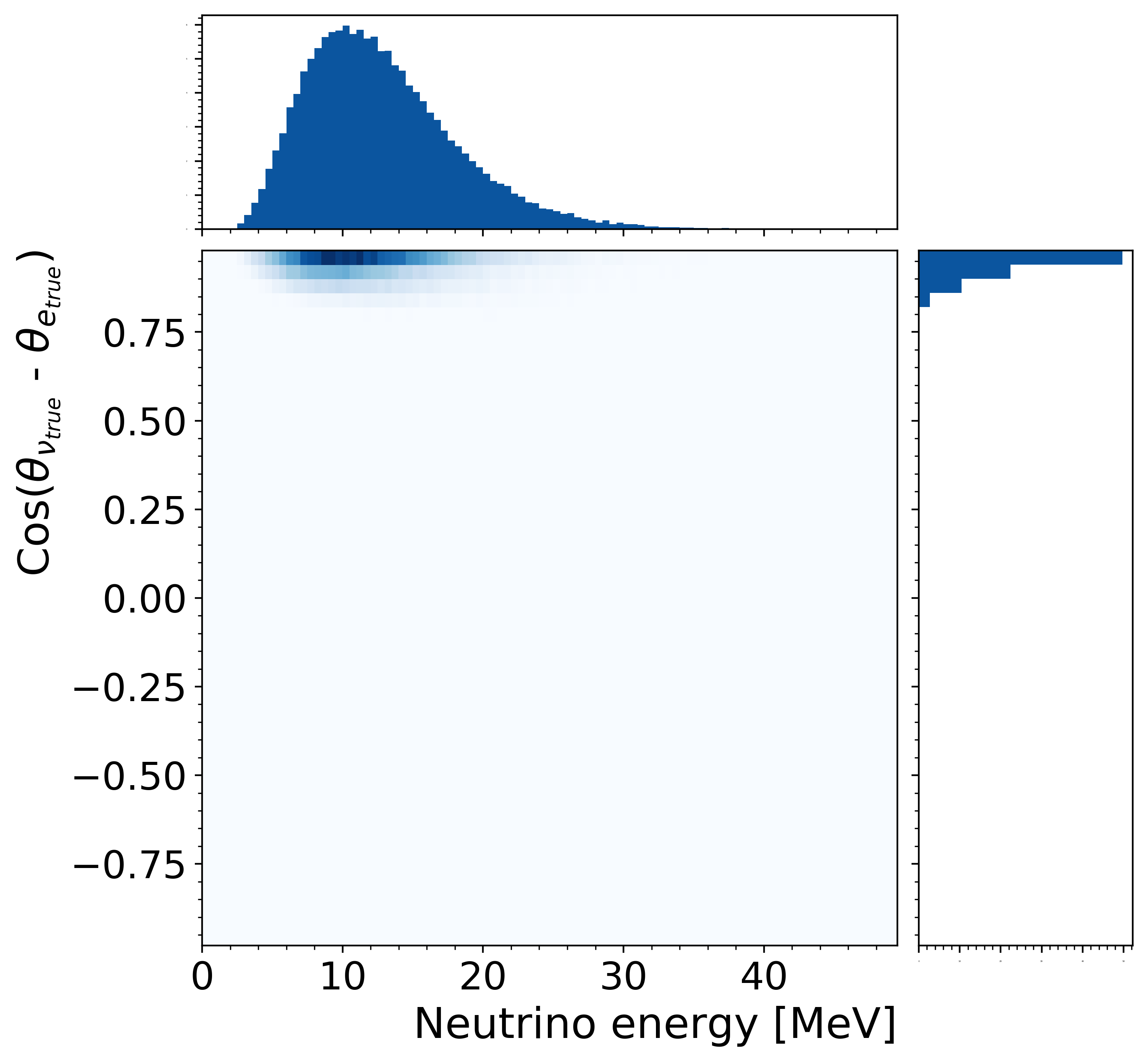}
    \caption{Histogram of the true neutrino energy as a function of cosine of the difference in the true neutrino angle ($\theta_{\nu_\text{true}}$) minus the true electron angle ($\theta_{e_\text{true}}$) for $\nu_e$~CC events (left) and for ${\nu_e\text{--}e^{-}}$~ES (right).  The $z$-axis color scale units are events per 0.5~MeV per 0.04.}
    \label{fig:truth_energy_direction}
\end{figure*}

For the analysis presented here, we focus on reconstructing the primary electron coming from electron neutrino charged-current ($\nu_e$~CC) interactions and from electron neutrino--electron elastic scattering (${\nu_e\text{--}e^{-}}$~ES). As can be seen in Fig.~\ref{fig:truth_energy_direction}, only the ES events preserve most of the progenitor neutrino directionality and can be used to perform neutrino pointing.

\begin{figure}[htb!]
  \centering
  \includegraphics[width=0.95\textwidth]{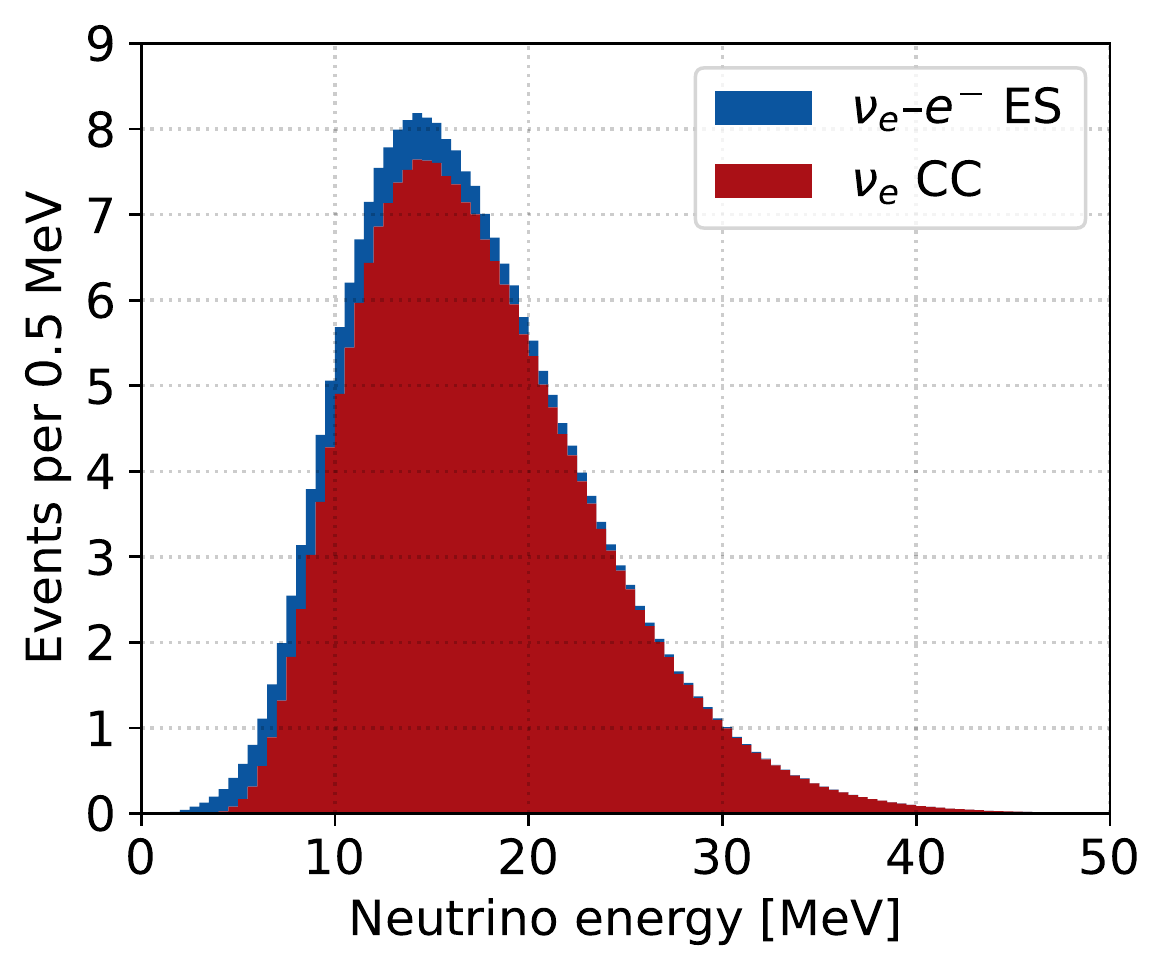}
  \caption{Stacked energy spectrum of interacted supernova neutrinos with 220 $\nu_e$~CC (red) and 19 ${\nu_e\text{--}e^{-}}$~ES events (blue) in 10 kton of liquid argon computed using {\tt SNOwGLoBES}.}
  \label{fig:neutrino-energy}
\end{figure}

\begin{table}[htb!]
    \centering
    \resizebox{0.925\textwidth}{!}{
    \setlength{\tabcolsep}{10pt}%
    \def\arraystretch{1.2}%
    \begin{tabular}{cc}
        \toprule
        Interaction channel                                           & Number of events \\
        \midrule
        $\nu_e + \ce{^{40}Ar} \rightarrow e^- + \ce{^{40}K^*}$        & 220              \\
        $\bar{\nu}_e + \ce{^{40}Ar} \rightarrow e^+ + \ce{^{40}Cl^*}$ & 5                \\ 
        $\nu_e + e^- \rightarrow \nu_e + e^-$                         & 19               \\
        $\bar{\nu}_e + e^- \rightarrow \bar{\nu}_e + e^-$             & 4                \\
        $\nu_\mu + e^- \rightarrow \nu_\mu + e^-$                     & 3                \\
        $\bar{\nu}_\mu + e^- \rightarrow \bar{\nu}_\mu + e^-$         & 2                \\
        $\nu_\tau + e^- \rightarrow \nu_\tau + e^-$                   & 3                \\
        $\bar{\nu}_\tau + e^- \rightarrow \bar{\nu}_\tau + e^-$       & 2                \\
        \bottomrule
    \end{tabular}
    }
    \caption{Event counts of $\nu_e$~CC, $\bar{\nu}_e$~CC, and ${\nu_X\text{--}e^{-}}$~ES interactions in 10~kton of liquid argon for a core-collapse supernova at 10~kpc computed with {\tt SNOwGLoBES}.}
    \label{tab:snb-event-counts}
\end{table}

\begin{figure*}[htb]
    \centering
    \subfloat {{\includegraphics[width=0.45\textwidth]{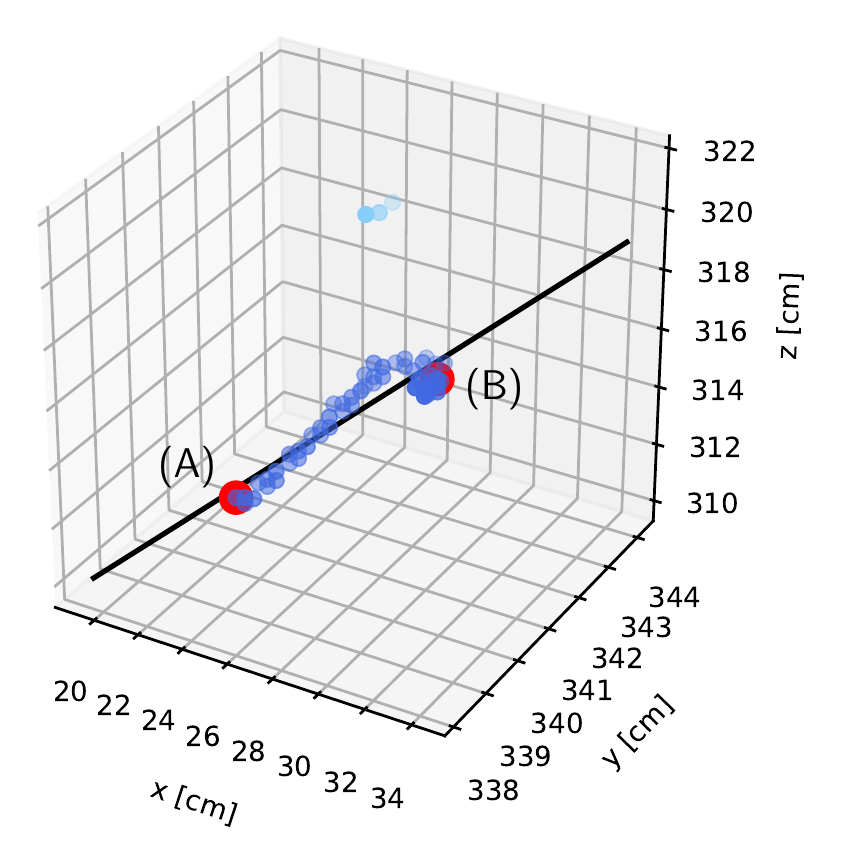} }}%
    \subfloat {{\includegraphics[width=0.50\textwidth]{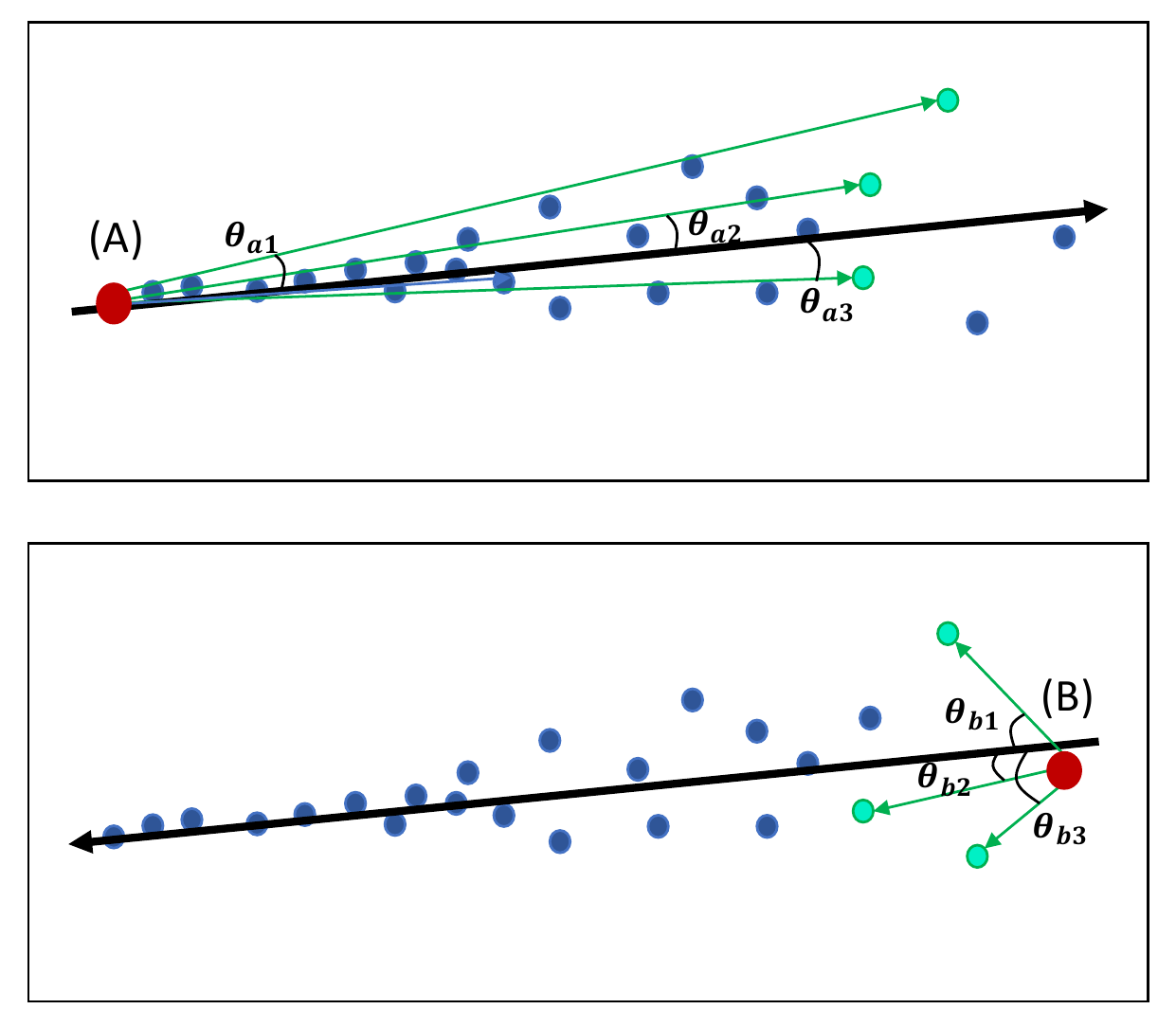} }}%
    \caption{Left: hits produced by the outgoing electron in an example supernova neutrino event. The black line is the linear RANSAC fit. The darker blue points and lighter blue points represent the inlier and outlier reset points identified by the RANSAC algorithm respectively. The two red dots (A) and (B) are the farthest reset points apart from each other among the inlier reset points. Right: 2D cartoon representation of inlier reset points. The two red dots are the left-most and right-most point among the inlier reset points. The black line represents the linear RANSAC fit performed to find the trajectory and the cyan points and green lines illustrate inlier reset points to identify the direction of travel of the electron.}%
    \label{fig:RANSACExample}%
\end{figure*}

To reconstruct the electron track for a candidate event, we first require that the event has at least 13 resets. Based on the analysis presented in Section~\ref{sec:EventID}, this ensures $<$1\% contamination from radiogenic backgrounds while still maintaining a relatively high signal detection efficiency. The collection of 3D reset positions are then analyzed using a random sample consensus (RANSAC) algorithm~\cite{RANSAC} in order to determine the resets that are along the main trunk of electron's path (inlier points) from resets coming from radiative processes from the electron interacting in the argon (outlier points). A graphical representation of this method can be seen on the left of Fig.~\ref{fig:RANSACExample} where inlier and outlier reset points are identified. Once so labeled, a RANSAC linear fit is performed on the inlier reset points and the reset points furthest in distance from each other are identified from the inlier reset points. The linear fit provides the axis for the direction determination, but there is an ambiguity as to which end of the line constitutes the starting reset point (SRP) and which end constitutes the ending reset point (ERP), as illustrated on the right of Fig.~\ref{fig:RANSACExample}. 

\begin{figure}[htb!]
    \centering
    \includegraphics[width=\textwidth]{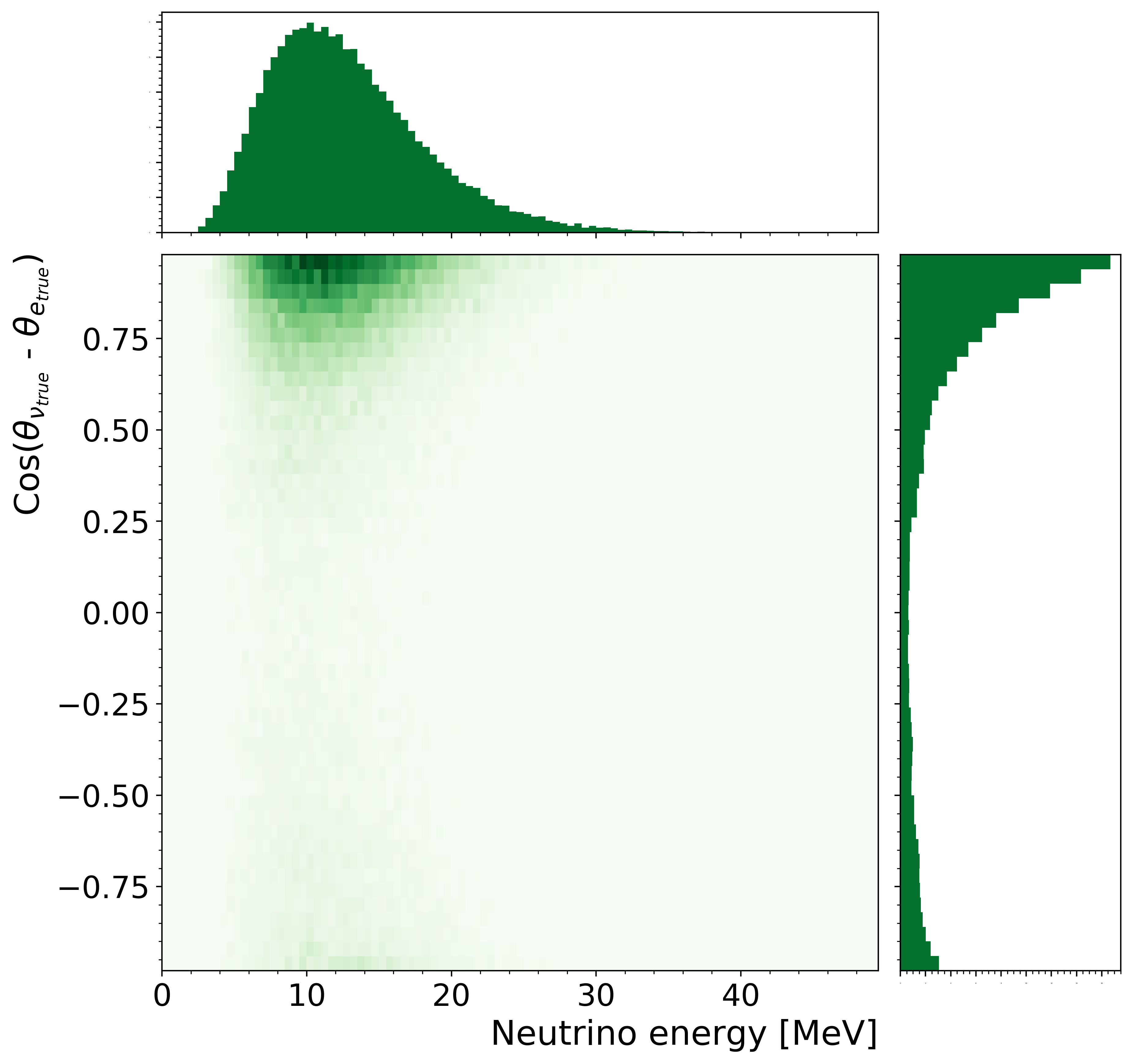}
    \caption{2D histogram for energy and angles between the neutrino's true momentum and the electron's reconstructed momentum after applying the directional correction for a sample of ${\nu_e\text{--}e^{-}}$~ES events.}
    \label{fig:es_true_reco}
\end{figure}

This ambiguity can be broken by taking a topological approach. As the primary electron travels in the argon, it will lose energy and experience larger scattering angles. Therefore, the spatial spread of the resets is a good indication of the directionality of the track. The right-hand side of Fig.~\ref{fig:RANSACExample} shows a 2D projection of the method used to determine which of the farthest inlier points should be considered as the SRP. The process begins by drawing a line between the two farthest reset points and then for each reset point within the inlier points, lines are drawn from the assumed SRP and ERP, and the cosine of the angles between each of those lines and the central line is calculated. The point with the larger sum of cosine of angles is taken to be the ERP. Fig.~\ref{fig:es_true_reco} shows the resulting difference between the true neutrino direction and the reconstructed neutrino direction as a function of neutrino energy for a sample of ${\nu_e\text{--}e^{-}}$~ES events after applying the correction for the ambiguity of the start point. A clear peak at ${\cos(\theta_{\nu_\text{true}} - \theta_{e_\text{reco}}) \approx 1}$ indicates that this reconstruction preserves the directionality with relatively high fidelity. 

\begin{figure*}[htb]
    \centering
    \includegraphics[width=0.39\textwidth]{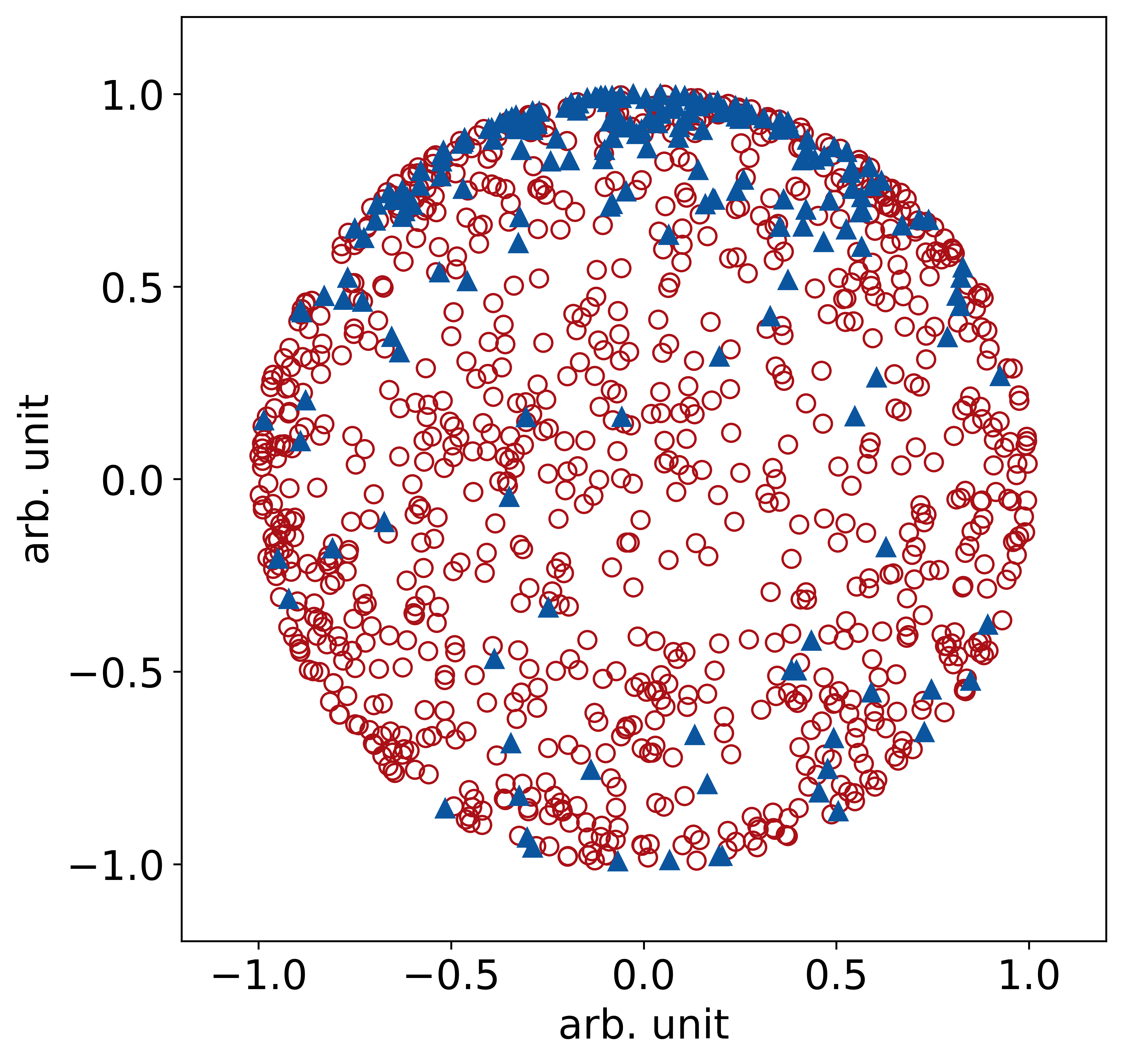} 
    \includegraphics[width=0.5\textwidth]{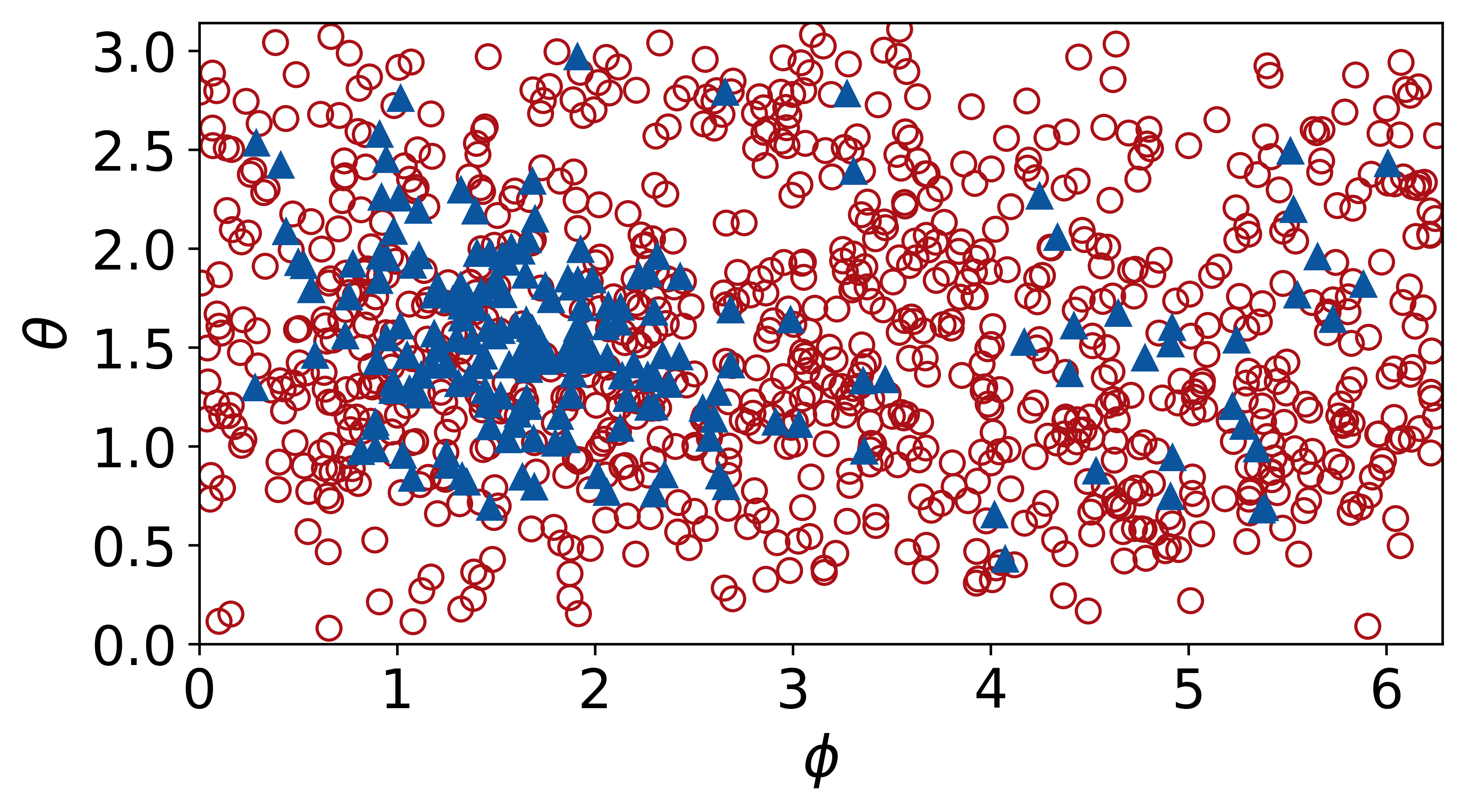}
    \caption{Representation of the direction vector generated for each neutrino interaction in a simulated supernova burst event shown on a unit sphere (left) and projected onto a $\theta$--$\phi$ plane. Solid blue triangles (\textcolor[HTML]{0B559F}{$\blacktriangle$}) represent events from ${\nu_e\text{--}e^{-}}$~ES interactions and hollow red circles (\textcolor[HTML]{AA1016}{$\circ$}) represent events from $\nu_e$~CC interactions.}
    \label{fig:bobtheblob}
\end{figure*}

With the directionality method established on an event-by-event basis, we now focus on how to use this information for an entire simulated supernova burst event. As mentioned before, we use the Garching benchmark model to simulate a typical burst topology observed in a DUNE 10-kton module, leading to 220 $\nu_e$~CC and 19 ${\nu_e\text{--}e^{-}}$~ES events (see Fig.~\ref{fig:neutrino-energy} and Table~\ref{tab:snb-event-counts}). Each neutrino interaction ($\nu_e$~CC and ${\nu_e\text{--}e^{-}}$~ES) generates a direction vector which we can project onto a unit sphere as is shown on the left of Fig.~\ref{fig:bobtheblob} (or projected onto a $\theta$--$\phi$ plane as seen on the right). Principal component analysis (PCA)~\cite{doi:10.1080/14786440109462720, 10.2307/2333955, jolliffe2002principal} is performed on this collection of points and a primary axis is chosen which penetrates the most populated area.

\begin{figure*}[htb]
    \centering
    \includegraphics[width=0.9\textwidth]{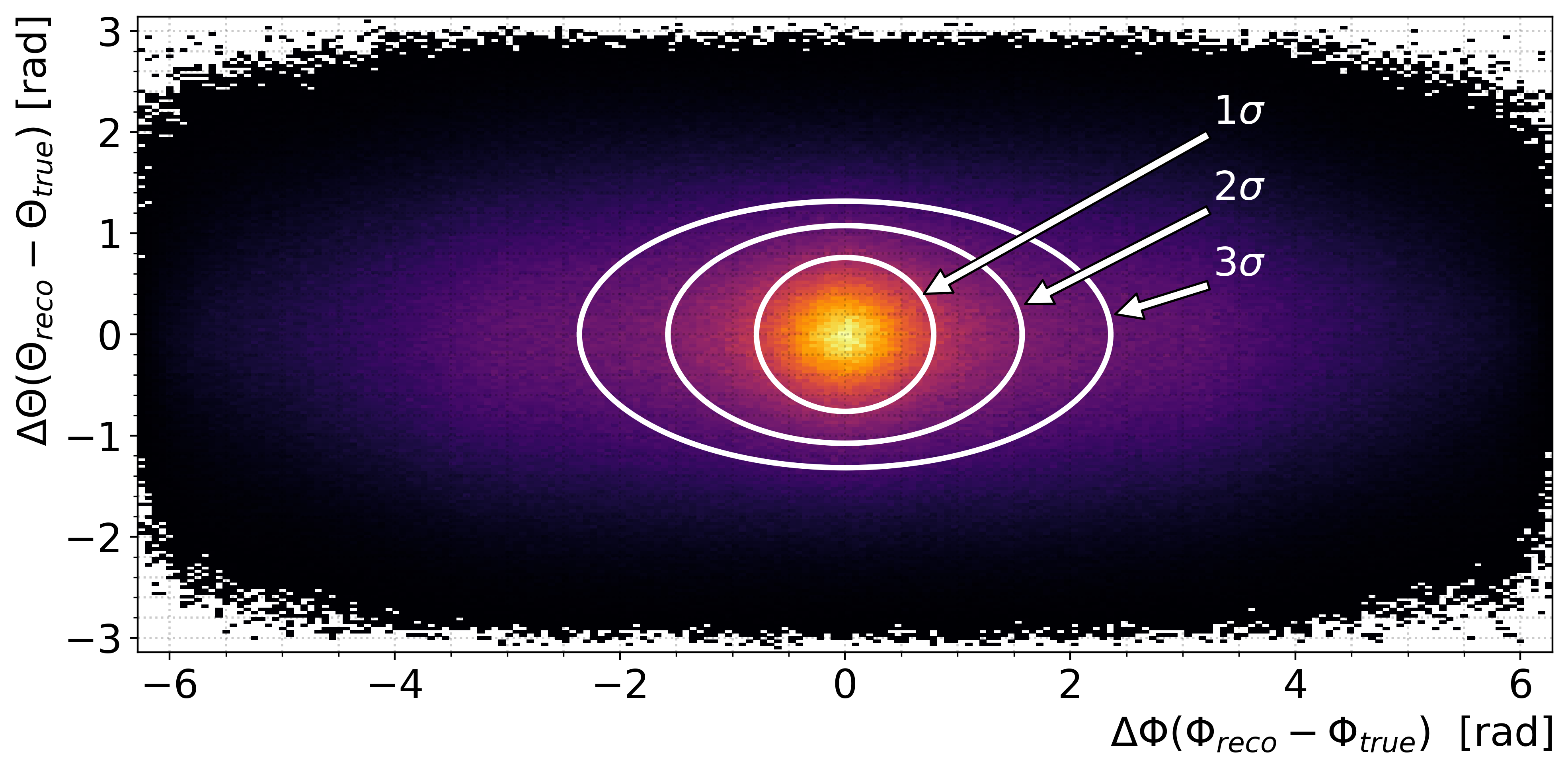} 
    \caption{Angular resolution for the supernova pointing in terms of $\theta$ and $\phi$. The $\sigma$ bands correspond to a 2D Gaussian fit to the peak to reduce the bias from the tails from poorly reconstructed events. A clear peak at  $\Delta \theta = \Delta \phi = 0$ can be observed. }
    \label{fig:angularAnalysis}
\end{figure*}

With the axis chosen, we can determine the direction of the supernova neutrinos using the same topological disambiguation method we used on an individual neutrino interaction, but now using all the points from the collection of neutrino interactions reconstructed. Using the most prominent point in this distribution to specify the direction from which the supernova burst occurred, we evaluate how well we reconstruct the supernova burst position by simulating 10,000 supernova burst directions and randomly distributing their origin. Analysis of this sample shows that for $\approx$80\% we correctly identify the direction of the supernova burst with a ${(\theta_{\text{reco}} - \theta_{\text{true}}) > 25^{\circ}}$ and the remaining $\approx$20\% have their direction incorrectly reported ``backwards'' with respect to the origin of the supernova burst. This restriction in the possible area of the sky a supernova burst search should be directed toward can provide a powerful tool for astronomical observations. In this analysis, we did not distinguish between ES and CC events, the latter of which can be identified by the presence of de-excitation photons. Identifying and removing these extraneous, non-pointing CC events would substantially improve the pointing analysis, and will be taken into account in future work. 

Fig.~\ref{fig:angularAnalysis} shows the angular resolution achievable with the methods described above. A clear peak at $\Delta \theta = \Delta \phi = 0$ can be seen. The $\phi$-projection is a Gaussian distribution while the $\theta$-projection has two additional shoulders around $-\pi$ and $\pi$ due to cases where the reconstructed direction points to the opposite of the true direction. The $1\sigma$, $2\sigma$, and $3\sigma$ contours of a 2D Gaussian fit to the difference between the true neutrino direction and the reconstructed neutrino direction are also shown. These results show that the 10~kpc supernova would be reconstructed within $\theta=33^{\circ}$ and $\phi=45^{\circ}$ at 1$\sigma$, and $\theta=99^{\circ}$ and $\phi=135^{\circ}$ at 3$\sigma$.

\section{Conclusions}\label{sec:Conclusion}
The opportunity to observe neutrinos emitted from the next core-collapse supernova offers a unique laboratory to test our understanding of particle physics under some of the most extreme conditions. These supernova produce a large number of neutrinos in the MeV energy range and a measurement of the timing and energy profiles can potentially provide answers to many astrophysical questions. As these neutrinos emerge promptly from a core-collapsing star, while the first observable electromagnetic signals may not manifest for hours to days later, the prompt identification of the neutrinos can be used to provide an early warning of an imminent visible supernova.

The Q-Pix detector concept presented in this paper provides a path to pixelated, kiloton-scale LArTPCs providing low-energy threshold detection that maximizes the physics potential while keeping data rates manageable and preserving the 3D information.

Using standard simulation tools for supernova neutrino interactions in LArTPCs, the Q-Pix readout can offer low-energy detection and reconstruction capabilities that have already been demonstrated with the conventional projective wire-based readout currently envisioned for DUNE. Even when we take into account estimated radiogenic backgrounds coming from the bulk liquid argon and detector material, which has not previously been considered in published work, the Q-Pix readout offers very high efficiency and purity event identification for individual supernova neutrino interactions and the ability to have a near-line, near real-time supernova trigger. Finally, due to the preservation of the 3D information afforded by a pixel-based readout, an angular resolution of 33$^{\circ}$ in $\theta$ direction, and 45$^{\circ}$ in $\phi$ direction for 1$\sigma$ can be achieved within a single 10-kton module.
The easily achievable physics reach provided by the Q-Pix readout, especially in the area of low-energy physics, makes compelling the further pursuit of this technology for kiloton-scale LArTPCs. Future work is currently underway to fully demonstrate the capabilities of this readout by examining beam, solar, atmospheric neutrinos as well as the capability to explore various beyond the Standard Model signatures.

\begin{acknowledgements}
This material is based upon work supported by
the U.S.\ Department of Energy, Office of Science, Office of
High Energy Physics Award No.~DE-0000253485 and No.~DE-SC0020065. R.~Guenette is also supported by the Alfred P.\ Sloan Foundation. S.~Kubota is also supported by the Ezoe Memorial Recruit and Masason Foundations. J.~B.~R.~Battat is supported by the Brachman Hoffman Fellowship through Wellesley College.
\end{acknowledgements}

\newpage 
\bibliographystyle{apsrev4-2}
\bibliography{QPIX_Supernova}

\end{document}